\documentclass[journal]{IEEEtran}

\usepackage[utf8]{inputenc}
\usepackage{booktabs}
\usepackage{framed} 
\usepackage{multicol} 
\usepackage{nomencl} 
\usepackage[boxed,ruled,commentsnumbered]{algorithm2e}
\usepackage{url}
\usepackage{soul}%
\usepackage{graphicx}

\usepackage{amsmath}
\DeclareMathOperator{\sign}{sign}
\usepackage{diagbox}
\usepackage{amssymb}
\usepackage{arydshln}
\usepackage{threeparttable}

\usepackage{algpseudocode}
\usepackage{subcaption}
\usepackage{yhmath}
\usepackage{multirow}
\usepackage{bm}
\usepackage{array}
\usepackage{epstopdf}
\usepackage{color}
\usepackage{stfloats}
\usepackage{indentfirst}
\usepackage{yhmath}
\usepackage{stfloats}
\usepackage{array} 
\usepackage{mathrsfs}
\newcolumntype{L}[1]{>{\raggedright\let\newline\\\arraybackslash\hspace{0pt}}m{#1}} 
\newcolumntype{C}[1]{>{\centering\let\newline\\\arraybackslash\hspace{0pt}}m{#1}} 
\newcolumntype{R}[1]{>{\raggedleft\let\newline\\\arraybackslash\hspace{0pt}}m{#1}}
\usepackage{geometry}
\newtheorem{remark}{Remark}[section]
\usepackage{tikz}
\usetikzlibrary{matrix, decorations.pathreplacing}

\ifCLASSINFOpdf
\else
\fi
\begin{document}

\title{Human-Machine Adaptive Shared Control for Safe Automated Driving under Automation Degradation}

\author{Chao Huang, Chen Lv,~\IEEEmembership{Senior Member, IEEE}, Peng Hang, Zhongxu Hu, Yang Xing
\IEEEcompsocitemizethanks{\IEEEcompsocthanksitem C. Huang, C. Lv,  P. Hang, Y. Xing and Z. Hu are with the School of Mechanical and Aerospace Engineering, Nanyang Technological University, 639798, Singapore. (E-mails: \{chao.huang, lyuchen, peng.hang, zhongxu.hu, xing.yang\}@ntu.edu.sg)}}

\maketitle

\begin{abstract}

In this paper, a human-machine adaptive shared control method is proposed for automated vehicles (AVs) under automation performance degradation. First, a novel risk assessment module is proposed to monitor  driving behaviour and evaluate  automation performance degradation for AVs. Then, an adaptive control authority allocation module is developed. In the event of any performance degradation detection, the allocated control authority of automation system is decreased based on the assessed risk to reduce the potential risk of vehicle motion. Consequently, the control authority allocated to the human driver is adaptively increased and thus requires more driver engagement in the control loop to compensate for the automation degradation and ensure the AV safety. Experimental validation is conducted under different driving scenarios. The testing results show that the proposed approach is able to effectively compensate for the performance degradation of vehicle automation through the human-machine adaptive shared control, ensuring the safety of automated driving.

\end{abstract}


\IEEEpeerreviewmaketitle

\section{Introduction} \label{intro}
\IEEEPARstart{I}{N} recent years, research institutes, automotive manufacturers, and IT companies have been making a great effort to  develop autonomous driving technology, since it has great potential to improve the safety and efficiency of transportation systems \cite{xu2019integrated, hang2020human}. Although fully autonomous driving is the ultimate goal, many technical problems remain to be solved and highly automated vehicles are considered to play a significant role before realizing fully autonomous driving
\cite{hoc2001towards}. Highly automated driving, which still retains the human driver in the control loop, presents an exciting new development in vehicle technology \cite{lv2017analysis,wang2020decision}. The approach allows the human driver and automation system to share the control authority and cooperatively operate the vehicle \cite{marcano2020review,huang2020reference}. This poses a new challenge, namely how to ensure a safe and smooth control allocation or transition between humans and automation during their shared control.

In a shared control algorithm design, one of the most critical issues is the control authority allocation. An improper authority allocation would lead to unsatisfactory driving performance and safety risks \cite{flemisch2012towards}. 
The control authority determines the amount of control effort provided by the automation and human driver \cite{benloucif2017new}. A lower level of control authority allocated to the automation would lead to little guidance and assistance provided to the human driver. However, a higher level of automation authority would result in the human driver's loss of domination \cite{nguyen2016driver}.

 Recent studies suggest that the allocation of the control authority should be adaptive to the dynamic states of the human driver and automation system, so that the driver and automation system have a complementary relationship that achieves better system performance and successful task completion \cite{vasconez2019design}. Ref \cite{flemisch2012towards} indicates that the control authority should be allocated based on the human driver's states \cite{wada2017shared}. Nevertheless, correctly identifying the human driver's states and performance, particularly under various complex situations, is still challenging \cite{erden2010human, xing2019driver}.

Although many studies in shared control were conducted  to improve drivers' control performance and vehicle safety, the proposed approaches usually rest on the assumption that automation systems are more reliable than drivers. Thus, the automation is usually expected to provide compensation to the human driver's undesirable actions. However,  the ability and maturity of autonomous driving technology are currently still limited. Many factors, such as complex traffic conditions, weather conditions, roadway types, and hardware and software issues, result in performance degradation and failure of vehicle automation. Based on different levels of severity, automation issues can be classified into two categories: the complete system failure and performance degradation  \cite{huang2018delta,8114322}.

In terms of the system failure, for example, if the power supply unit were to fail, all electronic components would be unable to function, which would lead to a catastrophic failure of the entire automation system. In this situation, an immediate handover of the control authority from automation to the human driver is required. However, in most occasions of real-world applications, even if some issues were to occur, the automation system would still be capable of working with degraded performance, rather than  completely failing. For example, the failure of the positioning logic  could result in an automated vehicle deviating from its expected lane. Furthermore, weather conditions, such as sun glare, rain or twilight, could lead to declined functionality of obstacle detection in an automated vehicle \cite{lv2017analysis}. Although the above automation degradation occurrences would affect the safe operation of AVs, it is still possible to mitigate such undesirable impacts by leveraging human assistance. During occurrences of automation degradation, the human driver is able to actively detect any potential driving risks and contribute more control effort accordingly, compensating  for the automation system’s degraded performance and ensuring the safety of the entire human-vehicle system \cite{favaro2018autonomous,dixit2016autonomous}. Nevertheless, the results of human assistance-based shared control algorithm for AVs under automation degradation are rarely reported.

Therefore, based on the above discussion, it is reasonable to design a shared control algorithm that considers the situation of automation system degradation. In such cases of declining automation performance, a human driver should be responsible for monitoring the situation and correcting the automation's undesirable behaviours by implementing his or her compensation control actions \cite{waschl2019control}. However, one challenge of implementing human-machine shared control under automation degradation is the risk assessment of driving and automation degradation \cite{lefevre2014survey}. One common approach is to use indices to measure driving risks, including the time to collision, time headway for longitudinal safety, the time-to-lane crossing and the variable rumble strip for lateral safety \cite{nodine2019indicators}. Another approach is to distinguish safe driving areas from unsafe areas and predict the possibility of entering an unsafe area based on the vehicle's dynamic states \cite{wang2016driving}.

\begin{figure*}[b!]
	\begin{center}
		{\includegraphics[width=0.8\textwidth]{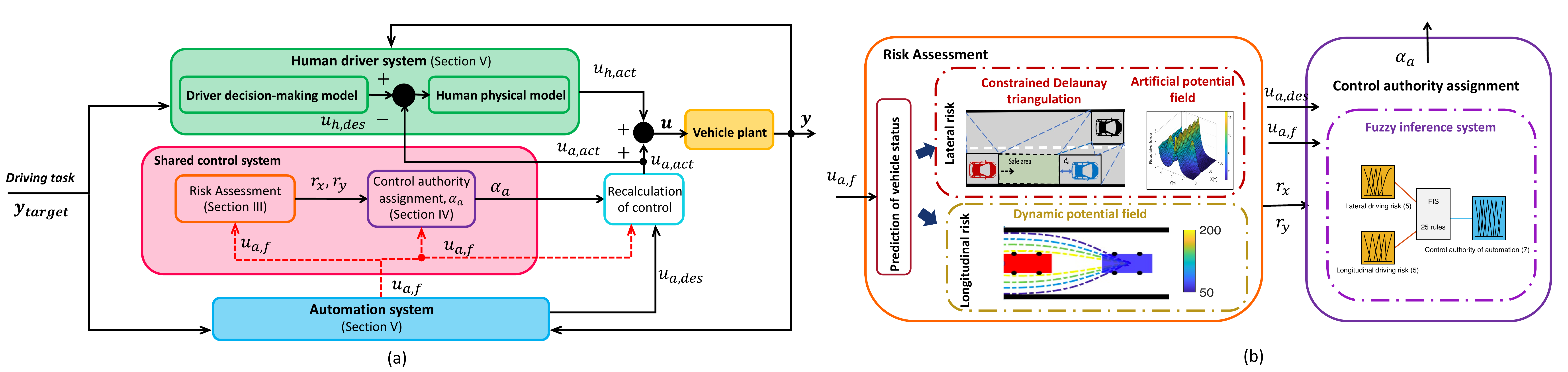}}
		\caption{Overview of the proposed shared control structure. (a) The high-level architecture of the algorithm; (b) Illustration of the risk assessment and control authority allocation modules.}
		\label{system_structure}
	\end{center}
\end{figure*}

In this paper, to further advance the shared control algorithm for AVs under automation degradation, we propose a novel approach. The proposed shared control architecture is composed of a \textit{risk assessment} module  and a \textit{control authority allocation} module. A unique feature of the proposed approach is that the system degradation is directly evaluated by a novel risk assessment module. Furthermore, this risk assessment module combines the above-mentioned two existing methods. It leverages the indices that are related to the safe area identification for measuring driving risks. This design allows for a continuous adaptation of the allocated control authority of the \textit{automation} according to the evaluated driving risk. To this end, the control authority, the control outputs of the degraded \textit{automation} system and the \textit{human driver}'s compensation actions are incorporated to construct the systematic shared control architecture. The main contribution is that the proposed approach offers a feasible and effective solution to the degraded \textit{automation} system of an AV through human-machine shared control. As a consequence, the \textit{human driver} would provide the compensation control effort and ensure the driving safety of the AV.

The remainder of the paper is as follows. The high-level structure of the shared control algorithm is introduced in Section \ref{systemstructe}.  The safety risk assessment for an AV in both longitudinal and lateral directions is designed in Section \ref{model}. The control authority allocation logic for  \textit{automation} is introduced in Section \ref{dsr}. \textit{Automation} and \textit{human driver} modelling are introduced in Section \ref{mp}.  The testing, validation and results are presented in Section \ref{sim}. Finally, Section \ref{con} concludes this paper.

\section{High-level Architecture of the Shared Control Algorithm}\label{systemstructe}

The following abbreviations are used throughout this paper:
\begin{itemize}[]
\item The suffixes $\bullet_a$ and $\bullet_h$ denote the variables issued from the \textit{automation} and \textit{human driver}, respectively.
\item The subscripts $act$ and $des$ are the abbreviations for the \textit{actual} and \textit{desired} control, respectively. They refer to the actual control vector imported to the vehicle and desired control vector computed by the \textit{automation} or \textit{human driver}. 
\item The subscript $f$  refers to the control output with a partial failure.
\end{itemize}

 The proposed architecture for human-machine shared control during vehicle automation degradation is depicted in Fig. \ref{system_structure}(a). For a specific task in automated driving , \textit{automation} is expected to control the vehicle to track a reference trajectory denoted by $\textbf{y}_{target}$. $\textbf{y}_{target}$ refers to the information of the reference trajectory, including the desired lateral coordinate of the desired lane centre $y^\oplus_{target}$ and the desired velocity $v_{target}$. In this paper, we mainly focus on the human-automation shared control algorithm design, while the $\textbf{y}_{target}$ is assumed to be available and its detailed generation method is not included. The desired control output, $u_{a,des}$, should be directly transmitted to the  actuator to execute an action (e.g., steering the front wheels). In this paper, we consider the performance degradation which that can caused by issues of software or hardware, such as a steering actuator or an electronic control unit. Under automation performance degradation, the \textit{human driver} is required to participate the control loop, providing compensation control, sharing the control authority with the automation and cooperatively operating the vehicle. The actual control vector applied to the vehicle plant, consisting of $u_{a,act}$ and $u_{h,act}$, is calculated as $\mathbf{u}=u_{h,act}+u_{a,act}$. In addition, the impact of the degraded automation performance on vehicle motion is evaluated by a risk assessment module. The corresponding upper bound of the control authority of \textit{automation} is managed by the control authority allocation module. The structure of the above components is shown in Fig. \ref{system_structure}(b) and will be detailed in Section \ref{model} and Section \ref{dsr}, respectively. The modelling of \textit{automation} and the \textit{human driver} are briefly discussed in Section \ref{mp}. Finally, the proposed shared control structure is expected to offer the \textit{human driver} an appropriate authority to compensate for the degraded automation performance and ensure the safety of the human-vehicle system. Further details are reported as follows.

\section{Safety Risk Assessment}\label{model}

In this section, we measure the safety risk based on the predicted vehicle  trajectory in both longitudinal and lateral directions. 
To ensure the lateral safety, we first use constrained Delaunay triangulation (CDT) to identify a safe area ($C$) for driving and then exhibit the boundaries of the safe area by using the artificial potential field (APF) model. We measure the potential field intensity of the predicted vehicle trajectory to evaluate the lateral driving risk. The closer the predicted trajectory is to the boundaries, the greater the potential field intensity is, and the higher the lateral risk would be. In terms of the longitudinal safety assessment, we use the dynamic Potential Field (DPF) to evaluate the possibility of a potential collision in the longitudinal direction by considering not only the vehicles' relative positions but also their velocity difference.

\subsection{Prediction of vehicle status}\label{pred}
In this section, we use the dynamic bicycle model and the \textit{constant turn rate and acceleration} (CTRA) model \cite{schubert2011empirical} to predict the vehicle's trajectory with the control input $u_{a,f}=[a_{a,f}\  \delta_{a,f}]^T$. The CTRA model has been widely used for prediction and tracking  \cite{benloucif2019cooperative}. It assumes that a circular path exists between two consecutive time steps of a vehicle, and the yaw rate and acceleration between the two time steps are constant.  For a degraded control vector $u_{a,f}$, the corresponding yaw rate $r_{ef}$, yaw angle $s_{ef}$ and lateral velocity $v_{x\_ef}$ can be calculated by using a dynamic bicycle model, which will be described in Section \ref{mp}. Here, subscript $e$ represents the \textit{ego} and subscript $f$ refers to the vehicle status computed by the degraded control inputs. Based on $r_{ef}$, $s_{ef}$ and $v_{x\_ef}$, the CTRA model is utilized to predict the vehicle's trajectory over a time horizon. The following equation describes the kinematics of the CTRA vehicle model:
\begin{equation}\label{prr}
    \xi(k+\tau_p)=\left\{
\begin{aligned}
x^\oplus_{ef}(k) & + \Delta x^\oplus_{ef}(\tau_p) \\
y^\oplus_{ef}(k) &+ \Delta y^\oplus_{ef}(\tau_p) \\ 
\psi_{ef}(k)&-\psi_{road}+\tau_pr_{ef}\\
v_{x\_ef}(k)&+a_{a,f}\tau_p\\
a_{a,f}&\\
r_{ef}(k)&\\
\end{aligned}
\right.
\end{equation}
with
\begin{equation*}
    \begin{aligned}
        \Delta x^\oplus_{ef}(\tau_p)&=\frac{1}{r^2_{ef}}[(v_{x\_ef}r_{ef}+a_{a,f}r_{ef}\tau_p)\sin(\psi_{ef}^\star(k)+\tau_pr_{ef})\\
        &+a_{a,f}\cos(\psi_{ef}^\star(k)+\tau_pr_{ef})\\
        &-v_{x\_ef}r_f\sin(\psi_{ef}^\star(k))-a_{a,f}\cos(\psi_{ef}^\star(k))]\\
            \end{aligned}
\end{equation*}
and
\begin{equation*}
    \begin{aligned}
        \Delta y^\oplus_{ef}(\tau_p)&=\frac{1}{r^2_{ef}}[(-v_{x\_ef}r_{ef}-a_{a,f}r_{ef}\tau_p)\cos(\psi_{ef}^\star(k)+\tau_pr_{ef})\\
        &+a_{a,f}\sin(\psi_{ef}^\star(k)+\tau_pr_{ef})\\
        &+v_{x\_ef}r_{ef}\cos(\psi_{ef}^\star(k))-a_{a,f}\sin(\psi_{ef}^\star(k))].
    \end{aligned}
\end{equation*}
where $\psi_{ef}^\star(k)=\psi_{ef}(k)-\psi_{road}$, and $\psi_{road}$ is the road heading angle. $\xi=[x^\oplus_{ef}\  y^\oplus_{ef}\  \psi_{ef}\  v_{x\_ef}\ a_{a,f}\   r_{ef}]^T$ is the state vector. $x^\oplus_{ef}$ and $y^\oplus_{ef}$ represent the vehicle's position. $\tau_p$ is the prediction horizon. Notably, a small prediction horizon may lead to a predicted trajectory that is close to the current position, which is unproductive for risk assessment. However, a large prediction horizon would result in over-shoots of the predicted trajectory and reduce the accuracy of the risk assessment. In this paper, we set $\tau_p=0.5s$. Once the predicted trajectory is generated, the predicted position $(x^\oplus_{ef}(\tau_p),y^\oplus_{ef}(\tau_p))$ is used to evaluate the driving risk. 

It should be noted that Eq. (\ref{prr}) is also used for the prediction of the system status of surrounding vehicles. For simplification, we omit subscript $f$, i.e., $x^\oplus_e(\tau_p)=x^\oplus_{ef}(\tau_p)$ and use $\Delta x^\oplus_{oe}=x^\oplus_o(\tau_p)-x^\oplus_e(\tau_p), o\in\{f,p,l\}$ to represent the relative distance between the \textit{ego} and surrounding vehicles in the longitudinal direction.  The subscript $o$ stands for the surrounding vehicles which includes $p$, $l$ and $f$. $p$, $l$ and $f$ stand for \textit{proceeding}, \textit{leader} and \textit{follower}, respectively (as shown in Fig. \ref{fig_safe}). In addition, we set $v_{j}$ as the abbreviation of $v_{x\_j}, j\in\{e,l,p,f\}$, and use $\Delta v_{eo}=v_e(\tau_p)-v_o(\tau_p),o \in \{p,l,f\}$ to describe the relative velocity in longitudinal direction. More details will be described in Section \ref{longr}.

\subsection{The lateral risk assessment}\label{inten}

The assessment of lateral driving risk is used to evaluate the possibility of system performance degradation, for example, caused by a faulty steering angle input. We describe the proposed lateral risk assessment algorithm, which is used for control authority allocation of \textit{automation}, as follows.

\subsubsection{Identification of the safe area}

\begin{figure}[t!]
	\begin{center}
		{\includegraphics[width=0.4\textwidth]{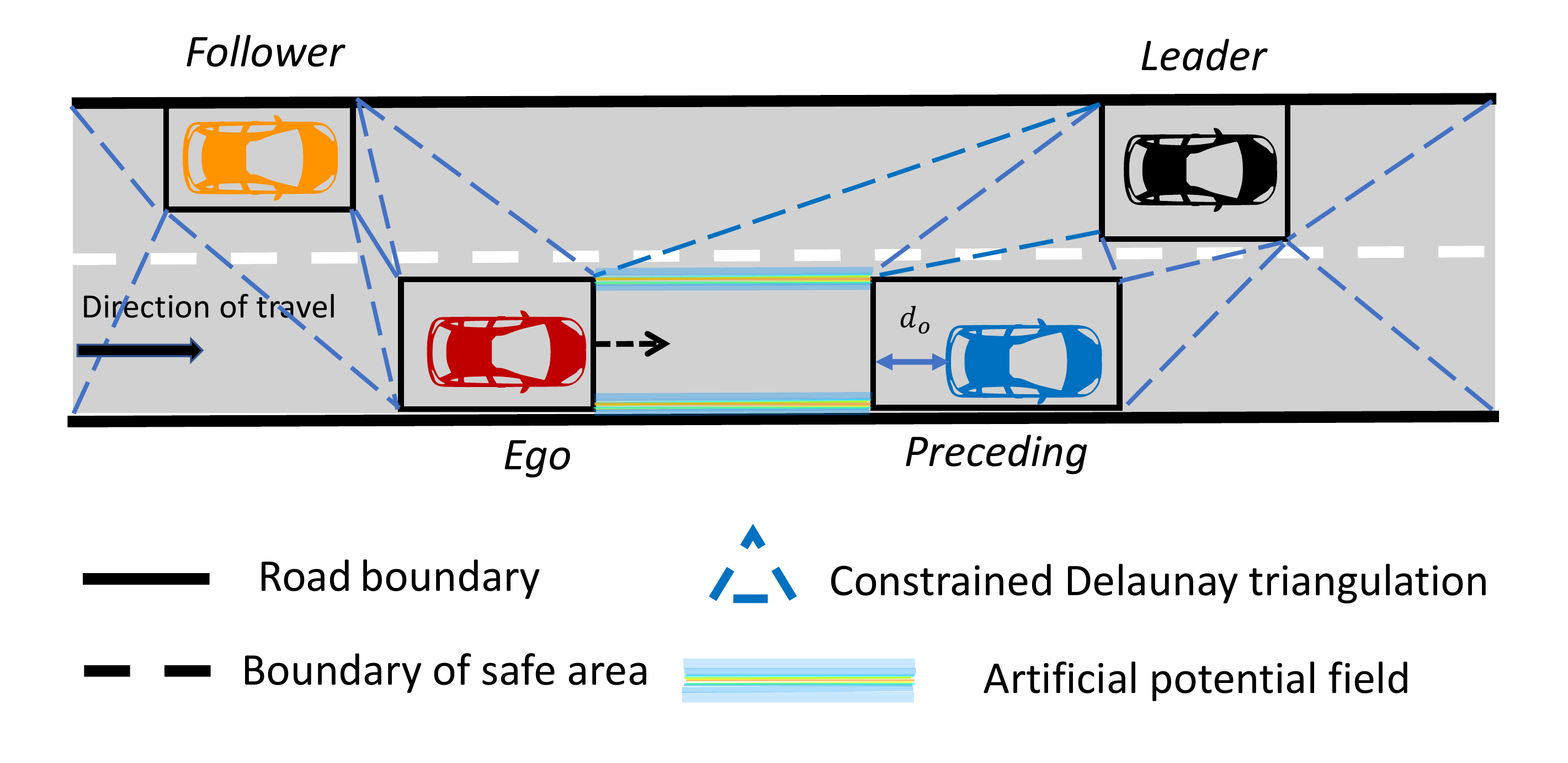}}
		\caption{Risk assessment in the lateral direction based on CDT and APF}
		\label{fig_safe}
	\end{center}
\end{figure}

We assume that the \textit{ego} vehicle is equipped with on-board sensors, including radars, LIDARs, cameras, and GPS/INS units. These on-board sensors enable the \textit{ego} vehicle to measure the position, speed and acceleration information of itself and the surrounding vehicles within a certain range. Therefore, the information of the driving environment can be obtained and further used for the generation of safe area. Given a driving region and the driving task (e.g. lane keeping in Fig. \ref{fig_safe}), the obstacles (moving vehicles including \textit{preceding}, \textit{leader} and \textit{follower}) and boundaries are described by line segments. The driving region can be partitioned into triangles. The safe area ($C$) can be considered a polygon that consists of a sequence of triangles connecting the starting point with the final region. The starting point is the \textit{ego}'s current position. The final region is defined as the position that is at a distance from the \textit{preceding}, where the distance $d_o$ is in the moving direction.

Based on the known information of the environment, the steps of identifying the safe area are summarized in TABLE \ref{algorithm1}. The boundaries of the safe area are considered nonovercome road boundaries. It should be noted that the boundaries consist of an \textit{upper} boundary and a \textit{lower} boundary in the lateral direction. More details can be found in \cite{huang2020reference}. The boundaries can be represented by the APF model and are discussed as follows.

\begin{table}[h!]

\caption{Algorithm for safe area ($C$) generation}\label{algorithm1}
\centering
  \begin{tabular}{ L{8cm}}
    \toprule
    1. Identify the starting point $S_p$ and final region $F_p$ and discretize the driving region into $N$ constrained Delaunay triangles;\\
    2. Find the triangle $T_s \in {\mathcal T}, 1\leq s \leq N, \mathcal{T}=\{T_1,T_2,\cdots,T_N\}$ to  which the starting point  $S_p$ belongs and the triangle $T_f \in {\mathcal T}$, $1\leq f \leq N, f\ne s$ to which the final point $F_p$  belongs.\\
    3. Construct an adjacent matrix $\bf{D}$$^{N\times N}$ to represent the relationship between any pair of triangles. $\bf{D}$$(i,j)=1$ represents that triangle $T_i$ and $T_j$ is adjacent. Otherwise, $\bf{D}$$(i,j)=0$\\
    4. Start with $T_s$, and based on the matrix $\bf{D}$, search for the triangle which is adjacent to $T_s$, repeat this process until $T_f$ is found.\\\bottomrule
    \end{tabular}

\end{table}

\subsubsection{The artificial potential field model}

The APF method was first proposed by Khatib in 1986 \cite{fan2020improved} and has been widely used for obstacle avoidance in robotics because it affords a simple mathematical analysis and entails less computation.  The following equation describes the calculation of the potential field of the boundaries:

\begin{equation} \label{ssss}
    P^r(x^\oplus_e,y^\oplus_e)=\left\{
\begin{array}{rcl}
&0, &  {r_b > d_c+\frac{v_w}{2}}\\
&\alpha_f \exp \left(-\frac{r_i^2}{\sigma_y^2} \right), &  Otherwise\\
\end{array} \right.
\end{equation}
where $r_b$ is the minimum distance from the vehicle's predicted position $(x^\oplus_e(\tau_p),y^\oplus_e(\tau_p))$ to the \textit{lower} boundary or the \textit{upper} boundary of the safe area. $v_w$ is the vehicle's width and $d_c$ is the safety margin. $\alpha_f$ is the coefficient to adjust the magnitude of the potential field value. $\sigma_y$ stands for the convergence coefficients along the direction of $Y$-axis, and it is set as half of the vehicle width, i.e., $\frac{v_w}{2}=1$. Based on Eq. (\ref{ssss}), it shows that when the vehicle's predicted positions remain away from the boundaries, the potential field approaches zero. Otherwise, the value of the potential field is inversely proportional to the distance. When \textit{ego}'s predicted positions reach the boundaries or move out of the safe area, the potential field value reaches its peak value, i.e., $\alpha_f$.  Therefore, we set the lateral risk as
\begin{equation}
    r_y=\max \left(0,\frac{P^r(x^\oplus_e,y^\oplus_e)}{\alpha_f}\right).
\end{equation}
It can be seen that  $r_y$ ranges from $0$ to $1$. The closer  $r_y$ is to $1$, the higher the lateral risk will be.

\begin{figure}[b!]
	\begin{center}
		{\includegraphics[width=0.31\textwidth]{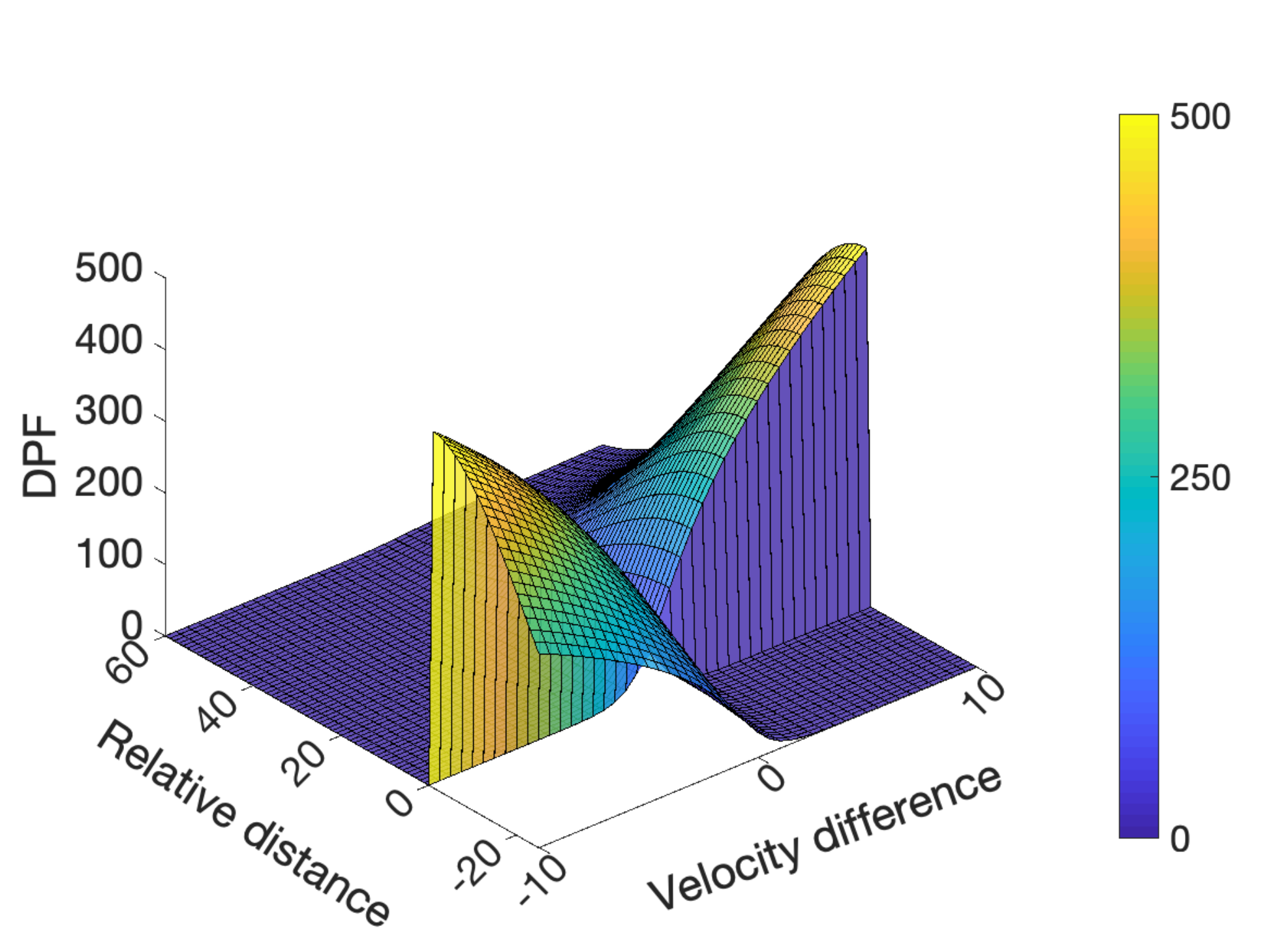}}
		\caption{Schematic diagram of the dynamic potential field designed based on the relative velocity and distance}
		\label{fig_dpf}
	\end{center}
\end{figure}
\begin{figure*}[b!]
    \centering
        \begin{subfigure}[b]{0.28\textwidth}
        \includegraphics[width=\textwidth]{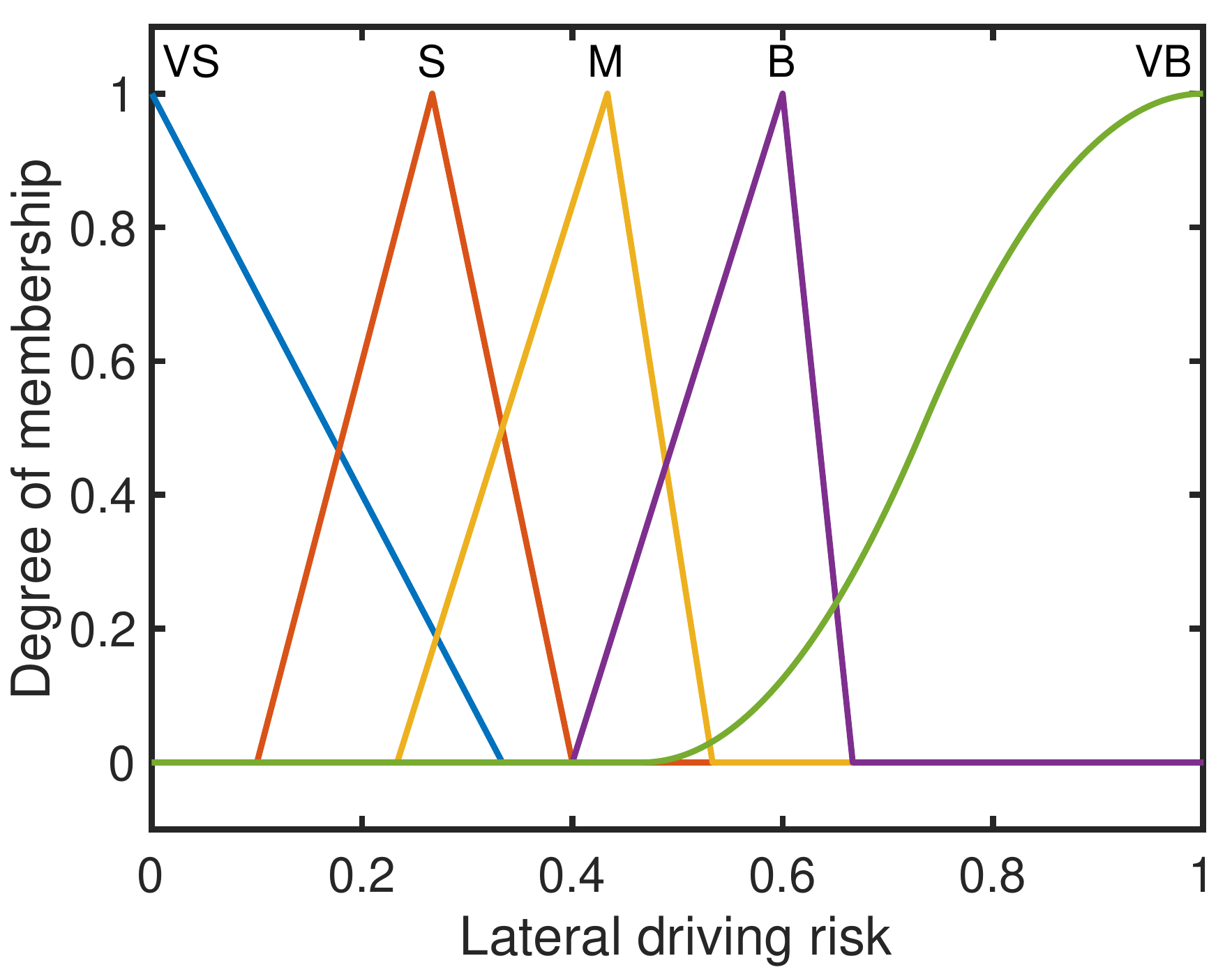}
        \caption{}\label{twolane_propose1}
    \end{subfigure}
        \begin{subfigure}[b]{0.28\textwidth}
        \includegraphics[width=\textwidth]{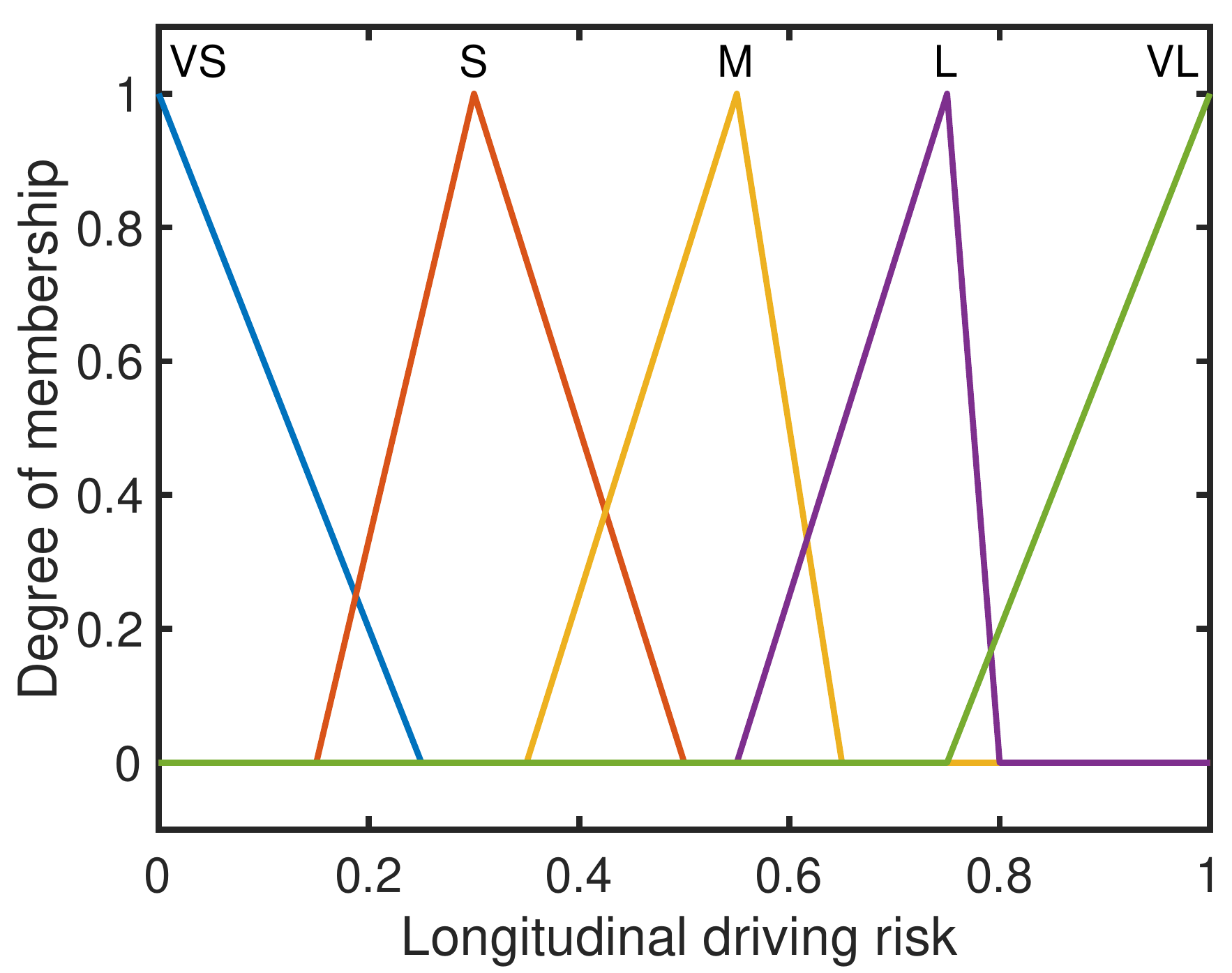}
        \caption{}\label{twolane_driver1}
    \end{subfigure}
            \begin{subfigure}[b]{0.292\textwidth}
        \includegraphics[width=\textwidth]{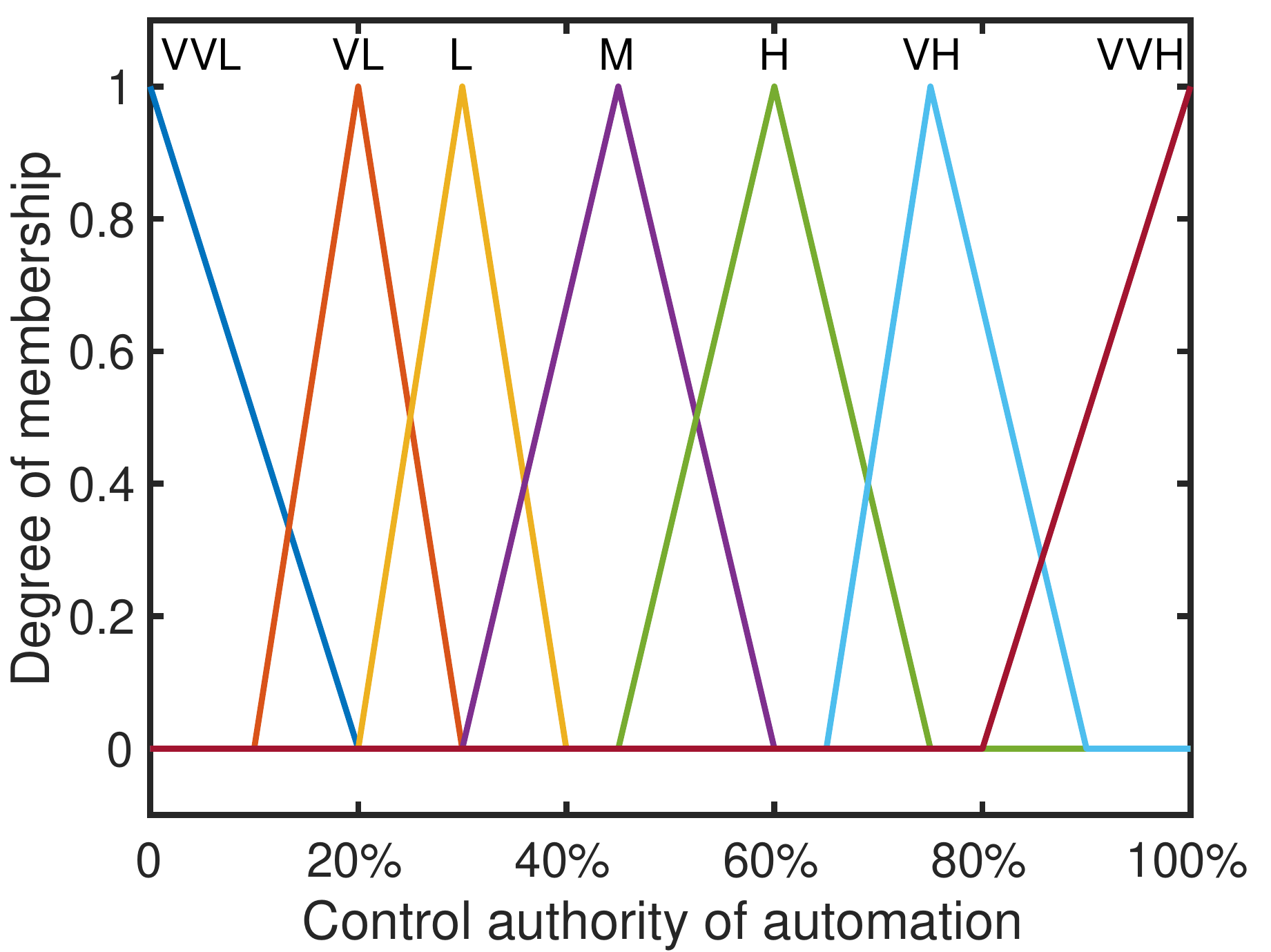}
        \caption{}\label{fig4}
    \end{subfigure}
    \caption{Membership functions: (a) the potential field; (b) vehicle's velocity; (c) the aggressiveness level. }\label{figfis}
\end{figure*}

\subsection{The longitudinal risk assessment}\label{longr}

In this subsection, we describe how to use DPF to evaluate the longitudinal driving risk. As we know, the possibility of  a collision occurring depends on the relative position and velocity between \textit{ego} and obstacles.  DPF ($U^d$) is defined as a function of the relative position and velocity of the \textit{ego} with respect to those of the obstacle, as shown in the following:

\begin{equation}
\label{sssa}
\begin{aligned}
    U^d(\Delta v_{eo},\Delta x_{oe})&= U_{eo} \cdot U_{rd}\\
    U_{eo}&=\alpha \frac{\exp[\eta(\Delta v_{eo})\cos(\theta)-I]}{2\pi I_0[\eta(\Delta v_{eo})]}\\
    U_{rd}&=A_f\cdot \exp\left[-\frac{(\Delta x^\oplus_{oe})^{2b}}{2 \sigma_x^{2b}}\right]\\
    \eta(\Delta v_{eo})&=\sign(\Delta v_{eo}) |\Delta v_{eo}|\\
    \end{aligned}
\end{equation}
with 
\begin{equation*}
    \theta=\left\{
\begin{aligned}
\pi, &\  x_e \leq x_o \\
0, &\ x_e > x_o \\
\end{aligned}
\right.
\ \ \ \ \ \ \ \ \ \  o \in \{p,l,f\}
\end{equation*}
where $U_{eo}$ stands for the potential field related to the relative velocity $\Delta v_{eo}=v_e(\tau_p)-v_o(\tau_p),o \in \{p,l,f\}$ and $U_{rd}$ represents the potential field associated with the longitudinal relative distance $\Delta x^\oplus_{oe}=x^\oplus_o(\tau_p)-x^\oplus_e(\tau_p), o \in \{p,l,f\}$.  The calculation of $\Delta v_{eo}$ and $\Delta x^\oplus_{oe}$ have been described in Section \ref{pred}. $\theta$ is the relative heading angle between two vehicles.  $\alpha$ and $A_f$ are the parameters correlated to the two potential fields, which are used to adjust the range of the potential field. $\sigma_x$ stands for the convergence coefficient along the direction of the $X$-axis. In our paper, we set the value of $\sigma_x$ as $2d_o$. $b$ is the coefficient for changing the shape. $I_0(\cdot)$ is the modified Bessel function of order 0. By introducing $U_{eo}$, the potential  field value can be drifted  to  the  direction $v_{eo}$.

Fig. \ref{fig_dpf} illustrates an example of the potential energy distribution based on the relative distance and velocity difference.  As shown on the right side of Fig. \ref{fig_dpf} ($\Delta v>0$), when the \textit{ego} moves faster than the  obstacle ahead (e.g., \textit{preceding}), the potential field value is significantly increased and reaches its peak value when the relative distance approaches zero. In contrast, the left side uf Fig. \ref{fig_dpf} describes the situation in which the velocity of the rear obstacle (e.g., \textit{follower}) is larger than that of \textit{ego}. The corresponding potential field value is also increased when the relative distance decreases. Moreover, when the \textit{ego} moves slower than the \textit{preceding} or the \textit{follower} moves slower than the \textit{ego}, the corresponding potential field value is no larger than 0 and indicates that the \textit{ego} is in a safe operating condition and the driving risk is low.

\begin{remark} It is reasonable to calculate the DPF of \textit{ego} with surrounding vehicles in a given driving manoeuvre. However, it should be noted that the longitudinal safety is also based on the driving task. For example, when the \textit{ego} is following the \textit{preceding}, the \textit{preceding}'s position and velocity would exert greater impact on the \textit{ego}'s longitudinal safety than that of the \textit{follower} and \textit{leader} in the adjacent lanes. Therefore, a weight should be associated with each DPF value that is adaptive to the driving task.
\end{remark}

Then, the following equation is used to calculate $r_x$:

\begin{equation} \label{r_long}
    r_x=w_f \frac{\max (0,U^d_{ef})}{U^d_{ef_{max}}}+w_l \frac{\max (0,U^d_{el})}{U^d_{el_{max}}}+w_p \frac{\max (0,U^d_{ep})}{U^d_{ep_{max}}}
\end{equation}
where $w_f$, $w_l$ and $w_p$ are the weights and satisfy $w_f+w_l+w_p=1$. In the lane-change driving task, the weights are set as $[0.8\ 0.1\ 0.1]$, while in the lane-keeping driving task, the weights are set as $[0.1\ 0.2\ 0.7]$. If the \textit{follower} or \textit{preceding} is not found, the corresponding weight will be set as 0. ${U^d_{eo}}, o\in \{f,l,p\}$ stands for the potential field between the \textit{ego} and \textit{follower}, and the \textit{leader} and \textit{preceding}, respectively. $U^d_{eo_{max}}, o\in \{f,l,p\}$ represents the corresponding maximum DPF. Under the assumption of a certain velocity difference, the maximum DPF is the potential field in the condition of a minimal safe distance. The corresponding calculation is given by
\begin{equation} \label{cc}
\begin{aligned}
U^d_{eo_{max}}&=\left\{
\begin{array}{rcl}
&I, &  U^d_{eo} \leq 0\\
&U^d(\Delta v_{eo},sign (x^\oplus_{oe})d^s_{\Delta_{eo}}), &  Otherwise\\
\end{array} \right.\\
    d^s_{eo}&=\frac{|v_e^2-v_o^2|}{2a_{max}}+\max(v_o,v_e)(t_r+\frac{t_i}{2})+d_o,\  o\in\{f,l,p\}.\\
    \end{aligned}
\end{equation}
where $d^s_{eo}$ represents the minimal safe distance. $a_{max}$ is the maximum deceleration for all vehicles. $t_r$ is the reaction time, and $t_i$ is the build-up time of deceleration. As the controlled vehicle is a highly automated vehicle, $t_i$ and $t_r$ are set as $0.1s$. $d_o=2\ [m]$ is the minimum clearance distance between the adjacent vehicles after they reach a same velocity. Based on Eq. (\ref{r_long}-\ref{cc}), it shows that for any pair $(e,o),o\in \{f,l,p\}$, if $U^d_{eo}$ is less than 0, the vehicle $o$ has little impact on \textit{ego}'s driving performance, and the term $\frac{\max (0,U^d_{eo})}{U^d_{eo_{max}}}$ is equal to 0. It is also indicated that $r_x$ is always less than 1. The closer $r_x$ is to 1, the higher the longitudinal risk will be.

\section{Adaptive Control Authority Allocation}\label{dsr}

We define $\alpha_a$ as the $\textit{maximum}$ control authority allocated to \textit{automation}. $\alpha_a$ is time varying and associated with the lateral and longitudinal driving risks. It is obvious that with the increase of the driving risk, the control authority of automation should be decreased, as well as $\alpha_a$. Here, we  utilize the Fuzzy inference system  (FIS) approach to build the relationship between the driving risk and $\alpha_a$. The advantage of utilizing a FIS is the capability of using $\textit{IF-and-THEN}$ rules to model knowledge through linguistic terms without employing precise quantitative analysis \cite{tavana2013fuzzy}.



First, we use linguistic terms to express the driving risk and control authority. We then use Triangular and S-shaped membership functions, which are the most commonly used membership functions, to fuzzify input and output variables:
\begin{enumerate}[]
\item Lateral driving risk. The fuzzy values of lateral driving risk can be categorized as very small (VS), small (S), medium (M), big (B) and very big (VB).
\item Longitudinal driving risk. The fuzzy values of longitudinal driving risk can be set as very small (VS), small (S), medium (M), large (L) and very large (VL).
\item Control authority of \textit{automation}. The fuzzy values of the FIS output are set as very very low (VVL), very low (VL), low (L), medium (M), high (H), very high (VH) and very very high (VVH).
\end{enumerate}
The corresponding membership functions are shown in Fig. \ref{figfis}. Second, we set fuzzy rules to describe the relationship between the system inputs and outputs. 
Last, we compute quantifiable results of the system output by using defuzzification methods. There are many different methods of defuzzification including adaptive integration (AI), center of area (COA), fuzzy clustering defuzzification (FCD), etc. We use the center of gravity defuzzification (COG) to achieve a crisp output.

\begin{figure}[t!]
	\begin{center}
		{\includegraphics[width=0.27\textwidth]{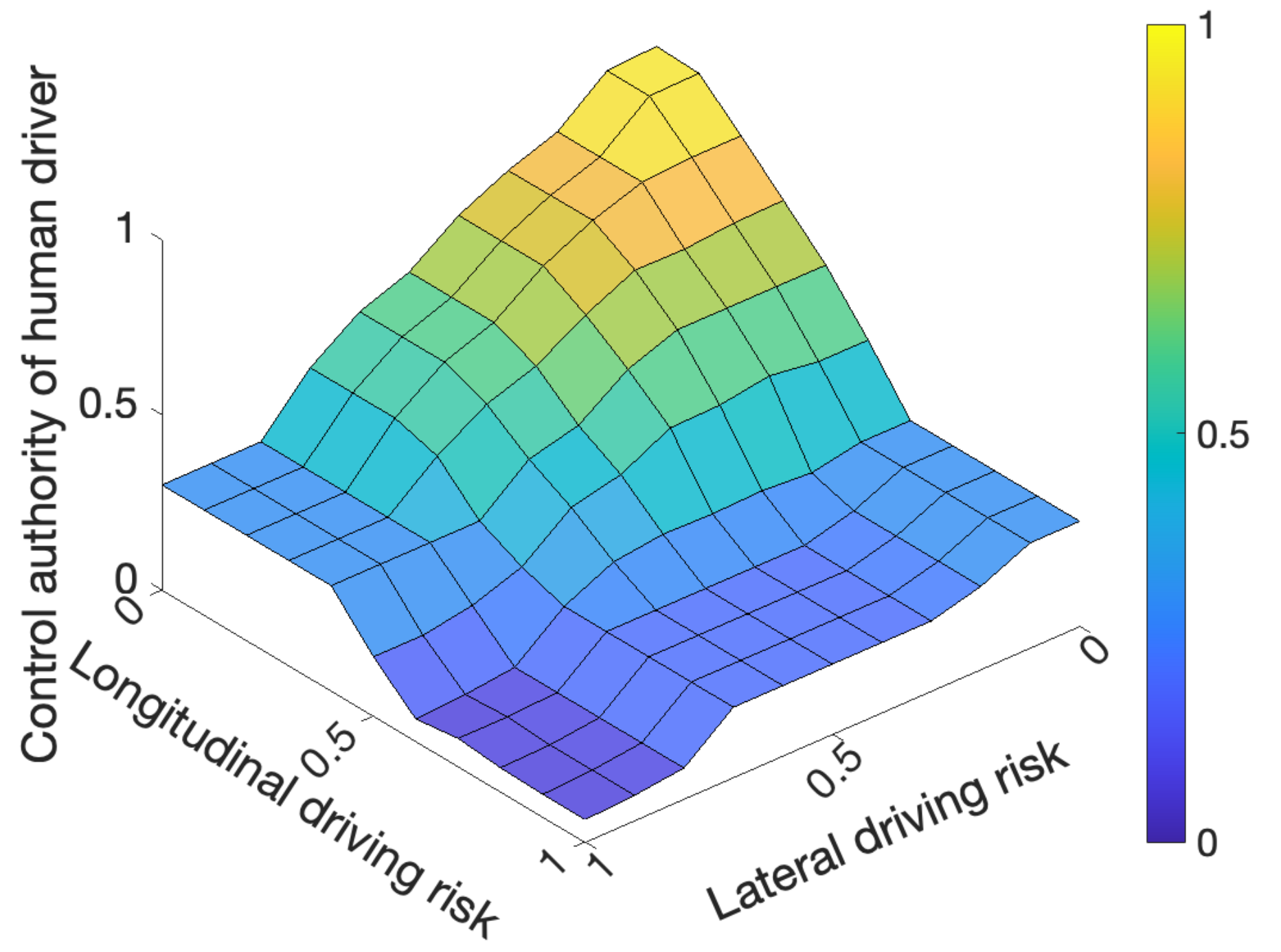}}
		\caption{Evolution of the values $\alpha_a$ according to longitudinal and lateral driving risks}
		\label{figdd_dm}
	\end{center}
\end{figure}

 The  output value $\alpha_r$ corresponding to the variation of the longitudinal and lateral risks over the universe of discourse is given in Fig. \ref{figdd_dm}.  It shows that when both longitudinal risk and lateral risk are low, the automation dominates the control of the vehicle. The corresponding $\alpha_a$ is very high ($\rightarrow 1$). However, the human driver is still responsible for monitoring the driving situation. The increase of either longitudinal risk or lateral risk indicates the occurrence of performance degradation and identifies the severity of the degradation. In this situation, the value of $\alpha_a$ will be decreased to limit the impact of $\textit{automation}$
on vehicle motion. Furthermore, the \textit{human driver} is required to assist the \textit{automation system} to take charge of the driving task with a higher value of allocated control authority.

 Considering $\alpha_a$, the calculation of the \textit{actual} control output $(u_{a,act})$ of \textit{automation} is given by
\begin{equation}
\label{actual_u}
    u_{a,act}(i)=\left\{
\begin{array}{rcl}
&u_{a,f}(i), \frac{u_{a,f}(i)}{ u_{a,des}(i)+\varepsilon} \simeq 1\\
&u_{a,f}(i),   0<\max \left(0, \frac{u_{a,f}(i)}{ u_{a,des}(i)+\varepsilon}\right)<\alpha_a <1\\
&\alpha_a \times u_{a,des}(i) ,  Otherwise.
\end{array} \right.\\
\end{equation}
for $i=1,\cdots,n$. $\varepsilon$ is an arbitrarily small positive quantity. In the normal operating condition, we have $u_{a,f}\simeq u_{a,des}$,  which meets the first condition of Eq. (\ref{actual_u}). Then, we obtain $u_{a,act}=u_{a,f}(u_{a,des})$. In the degraded condition, when the affected $\textit{automation}$ system has a small impact on vehicle motion, we have $u_{a,act}=u_{a,f}(<u_{a,des})$, which is described by the second term of Eq. (\ref{actual_u}). The \textit{human driver} is expected to provide the rest of the required control effort to compensate for degraded \textit{automation}. However, if $u_{a,f}$ is much larger than $u_{a,des}$ or they have different signs, we set  $u_{a,act}=\alpha_a \times u_{a,des}$. In this way, $u_{a,act}$ is always no larger than $u_{a,des}$. Therefore, the impact of the degraded $\textit{automation}$ system is limited.

\section{Automation and Driver System Modelling}\label{mp}

In this section, we first use a CDT-APF based MPC controller to model \textit{automation} in order to generate the desired control actions, i.e., $u_{a,des}$. Then, we introduce the driver decision-making model and human behavioural model.
\subsection{Modelling of automation system}
A dynamic vehicle model with three degrees of freedom is adopted for both controller design and  plant simulation, i.e.,

\begin{equation}
    \label{equ1}
    \begin{aligned}
    \dot{v}_x&=rv_y+a_{a,des},\\
    \dot{v}_y&=-rv_x+\frac{2}{m}(F_{cf}\cos \delta_{a,des}+F_{cr}),\\
    \dot{r}&=\frac{2}{I_z}(l_fF_{cf}-l_rF_{cr}),\\
    \dot{x}^\oplus&=v_x\cos \psi -v_y \sin \psi,\\
    \dot{y}^\oplus&=v_x \sin \psi +v_y \cos \psi.
    \end{aligned}
\end{equation}
where $x=[x^\oplus,y^\oplus,\psi,v_x,v_y,r]$ denotes the vehicle's status. $(x^\oplus,y^\oplus)$ denotes the vehicle's position in the body frame $(X,Y)$. $v_x$ and $v_y$ denote the longitudinal and lateral speeds of the centre of mass, respectively. $r=\dot{\psi}$ is the yaw rate and $\psi$ is the inertial heading angle. The system inputs are $u_{a,des}=[ a_{a,des}\ \delta_{a,des}]^T$ where $\delta_{a,des}$ is the front steering wheel angle and $a_{a,des}$ denotes the longitudinal acceleration. The definition and value selection of the key parameters of the vehicle model are listed in TABLE \ref{table:valuevehicledynamics}. $F_{cf}$ and $F_{cr}$ denote the lateral tyre forces at the front and rear wheels, respectively. 
$F_{ci}, (i \in \{f,r\})$ can be calculated by $F_{ci}=-C_{\alpha i}\alpha_i, (i \in \{f,r\}$. $C_{\alpha i} (i={f,r})$ denotes the stiffness coefficients for the front and rear tyres. $\alpha_i (i=f,r)$ denotes the tyre slip angle, which can be expressed as $\alpha_f \approx \frac{v_y+l_fr}{v_x}-\delta_{a,des}$ and $\alpha_r \approx \frac{v_y-l_rr}{v_x}$, respectively.

\begin{table}[h]
	\begin{center}
		\caption{Vehicle Parameters  }\label{table:valuevehicledynamics}
		\begin{tabular}{|c|c|c|}
			\hline
			\textbf{Parameter}  & \textbf{Description} & \textbf{Value} \\\hline
			$l_f$& Distances from CM to front axle & 1.21$m$ \\\hline
	                 $l_r$& Distance from CM to rear axle &1.05 $m$ \\\hline
		$m$&Vehicle mass & 2000 $kg$\\\hline
			$I_z$& Vehicle's moment of inertia &1300 $kg \cdot m^2$\\\hline
		\end{tabular}
	\end{center}
\end{table}

The overall vehicle system dynamics model is formulated as:
\begin{equation}
\label{system}
\begin{aligned}
    x(k+1)&=f^d(x(k),u(k))\\
    \mathbf{y}(k)&=p^r(x^\oplus(k),y^\oplus(k))
    \end{aligned}
    \end{equation}
where $\mathbf{y}(k)$ is the potential field value associated with the position coordinates based on CDT and APF. Based on the internal model built in Eq. (\ref{system}), the generation of desired control actions can be transferred to solve the following optimization problem:
\begin{equation}
\begin{aligned}\label{opt}
\min_{\mathcal{U}_a,\epsilon} \sum_{i=1}^{H_p} & \left \{  \Arrowvert  \mathbf{y}(k+i|k) \Arrowvert^2_R+\Arrowvert  y^\oplus(k+i|k)-y_{target} \Arrowvert^2_W \right.\\
&\left.+\Arrowvert  v(k+i|k)-v_{target} \Arrowvert^2_H+\Arrowvert \frac{dr(k+i)}{dt} \Arrowvert^2_N\right\} \\
    &\ \ \ \ \ \ \ \ \ \ \  \ \  \ \ \ \ \ \ \ \ \ \ \ \ \ +\sum_{i=1}^{H_c} \Arrowvert  u_{a,des}(k+i|k) \Arrowvert^2_Q \\
\end{aligned}
\end{equation}
 subject to
 
\begin{equation}\nonumber
\label{equ17}
\begin{aligned}
 \  &x(k+i+1|k)=f^d(x(k+i|k),u(k+i|k)),\\ \ 
    &\ \ \ \ \ \ u_{min} \leq u_{a,des}(k+i)\leq u_{max},\ i=0,\cdots,H_c-1\\
    &\ \ \ \ \ \ u(k-1|k)=u(k-1)\\
    &\ \ \ \ \ \ x(k|k)=x(k) 
    \end{aligned}
\end{equation}where $k$ denotes the current time, and $x(k+i|k)$ denotes the predicted state at time $k+i$ obtained by applying control sequence $\mathcal{U}_a = [u_{a,des}(k),\cdots, u_{a,des}(k+i)]$ to the system (\ref{system}). There are five objectives of the optimization problem. The first one is to minimize the potential field value. The second and the third objectives are to make the \textit{ego} drive consistently with the \textit{preceding} in the longitudinal direction and reduce their relative speed. The fourth term aims to minimize the yaw acceleration throughout the manoeuvre, ensuring vehicle stability. The final term reflects the consideration of the size of $u_{a,des}(k+i|k)$. $R$, $W$, $H$, $N$ and $Q$ are the weighting matrices for each objective. $H_c$ is the control horizon, while $H_p$ is the prediction horizon with $H_p \geq H_c$. In addition, the control inputs are subject to the physical limitations of the vehicles:
\begin{equation}
\begin{aligned}
\label{steer}
    \delta_{amin} \leq \delta_{a,des} \leq \delta_{amax}\\
    a_{amin} \leq a_{a,des} \leq a_{amax},
    \end{aligned}
\end{equation}
where $\delta_{amin}$, $\delta_{amax}$, $a_{amin}$ and $a_{amax}$  denote the saturation level of $\delta_{a,des}$ and $a_{a,des}$, respectively. The  above  optimization  problem  can  be  effectively  solved  by using the function of  \textit{fmincon} in MATLAB.


\subsection{Human driver system modelling}

The modelling of human driver is composed of a driver decision-making model and human behavioural model. The former model is used to generate the \textit{desired} control actions of a \textit{human driver} based on the driving task and environment information. The latter model describes the dynamic properties of a human driver's neuromuscular dynamics when performing a driving task. The generation of the \textit{desired} control action of the  \textit{human driver}, $u_{h,des}=[a_{h,des}\ \delta_{h,des}]^T$, is similar to Eq. (\ref{opt}), i.e. \cite{ji2018shared},

\begin{equation}
\begin{aligned}\label{opt1}
&\min_{\mathcal{U}_h,\epsilon} \sum_{i=1}^{H_p} \left\{ \Arrowvert  y^\oplus(k+i|k)-y_{target} \Arrowvert^2_W+\Arrowvert  v(k+i|k)-v_{target} \Arrowvert^2_H \right.\\
    &\ \ \ \ \ \ \ \ \ \ \ \ \ \ \ \ \ \ \ \left.+\Arrowvert \frac{dr(k+i|k)}{dt} \Arrowvert^2_N \right\}+ \sum_{i=1}^{H_c} \Arrowvert  u_{h,des}(k+i|k) \Arrowvert^2_Q\\
    &s.t.\  \  x(k+i+1|k)=f^d(x(k+i|k),u(k+i|k)),\\ 
    &\ \ \ \ \ \ u_{min} \leq u_{h,des}(k+i)\leq u_{max},\ i=0,\cdots,H_c-1\\
    &\ \ \ \ \ \ (x^\oplus(k+i|k), y^\oplus(k+i|k)) \in C, i=1,\cdots,H_p\\
    &\ \ \ \ \ \ u(k-1|k)=u(k-1)\\
    &\ \ \ \ \ \ x(k|k)=x(k) 
\end{aligned}
\end{equation}
The optimization objectives take the vehicle's lateral motion, vehicle velocity, yaw acceleration and size of the control action into account. The vehicle dynamics model and the design parameters, including $H,R,H_p$ are set as the same values as those in Eq. (\ref{opt}). In addition, the vehicle's predicted positions over the prediction horizon ($H_p$) should always belong to the safe area $C$. The output $u_{h,des}$ feeds the human driver model to generate the \textit{actual} control output, $u_{h,act}$. The model representing human driver's behaviour can be described as \cite{na2019modelling}:
\begin{equation} \label{driverm}
 u_{h,act}=\frac{K_h}{T_hs+1} (u_{h,des}-u_{a,act})
\end{equation}
where $K_h$ is the proportional gain, $T_h$ is time lag/delay of the driver's response, and $s$ is the Laplace operator. The validation of parameters $K_h$ and $T_h$ will be given in Section \ref{valid}.

\section{Experimental Testing, Validation and Results}\label{sim}

\begin{figure}[h!]
	\begin{center}
		{\includegraphics[width=0.32\textwidth]{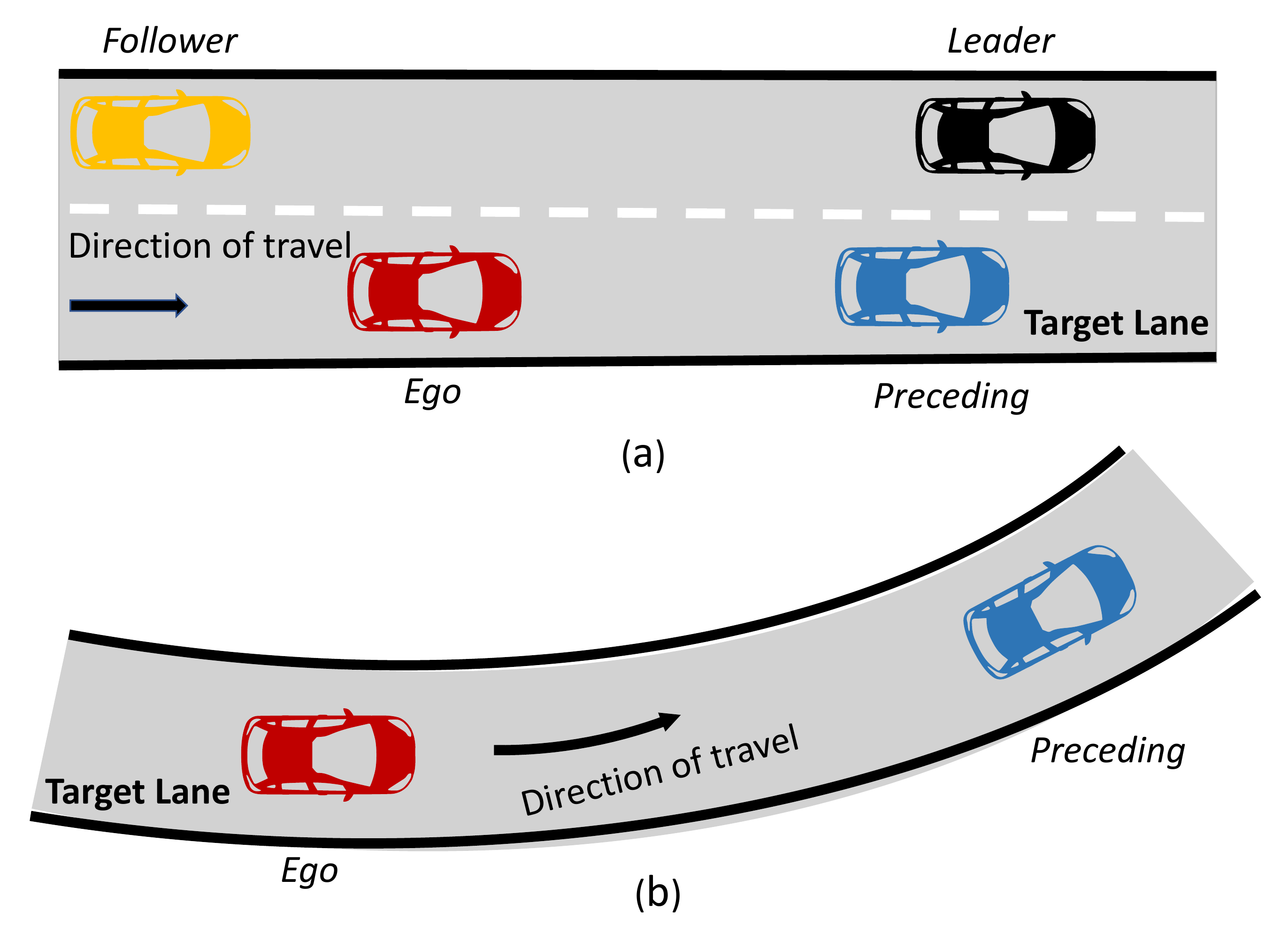}}
		\caption{Test cases: (a) case 1: a two-lane highway and (b) case 2: a curvy road  }
		\label{figdsfs_dm}
	\end{center}
\end{figure}

This section aims to verify the feasibility and effectiveness of the proposed shared control approach in a specific driving scenario of lane keeping. As shown in Fig. \ref{figdsfs_dm}, two different testing cases are carried out.  Case 1 is a two-lane scenario and Case 2 is a scenario of a curvy road.

 \begin{figure}[b!]
	\begin{center}
		{\includegraphics[width=0.35\textwidth]{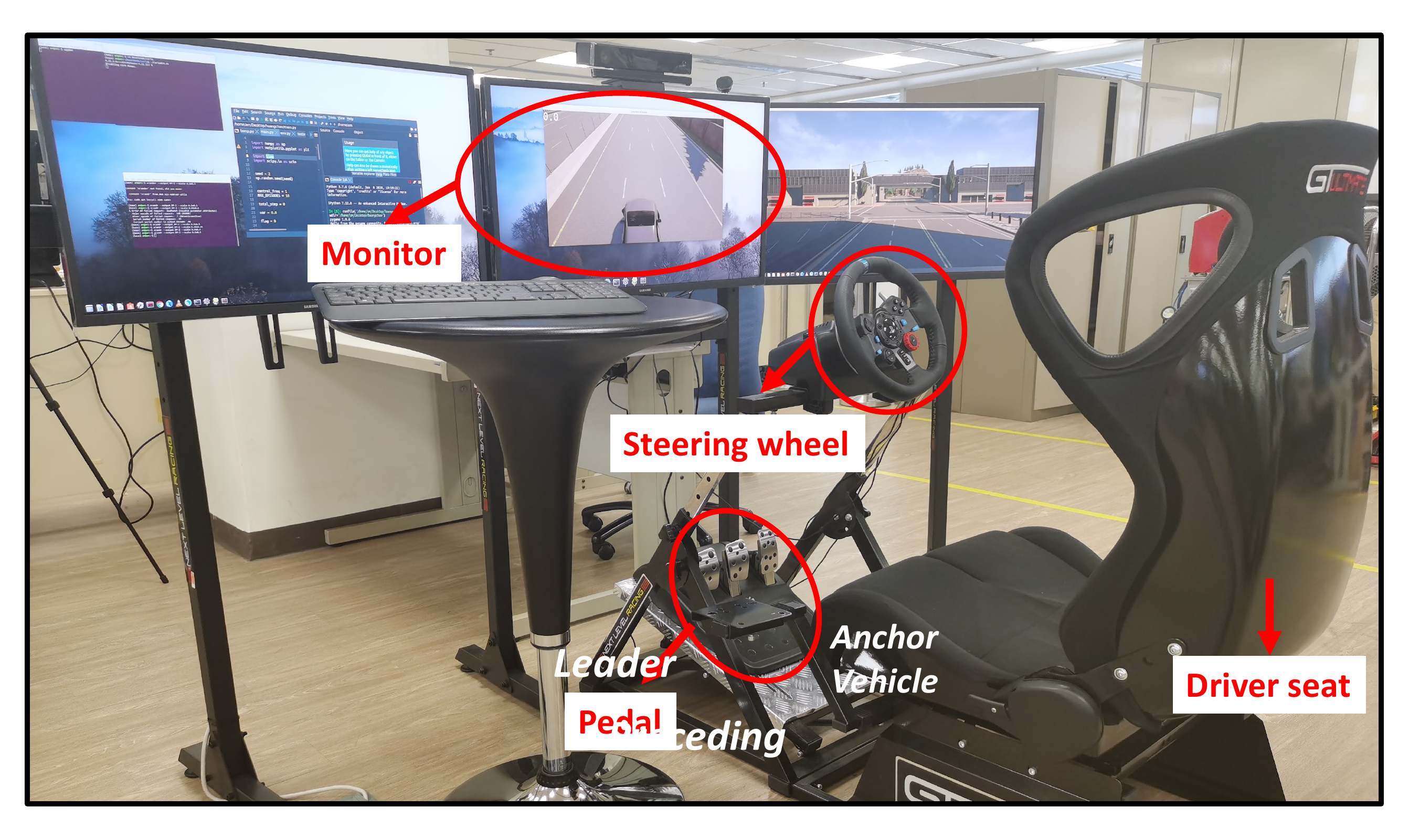}}
		\caption{Experimental platform with a CARLA simulation environment}  \label{expp}
	\end{center}
\end{figure}

\subsection{Driver model validation}\label{valid}
In this subsection, the parameter identification of the driver model is conducted. Based on Eq. (\ref{driverm}), the parameter vector $[K_h\ T_h]$ can be identified in the MATLAB \textit{system identification toolbox}. A total of 5 drivers (4 males, 1 female, mean age at 25.5 years) participated in the experiment.  Each participant holds a valid driver’s licence. The  experiment  was  conducted  using  a  driving  platform with  the CARLA  simulation environment,  as  shown  in  Fig. \ref{expp}. The  platform consists of a computer equipped with an NVIDIA GTX 2080 Super GPU, three joint head-up monitors, a Logitech G29 steering wheel suit, and a driver seat.  All the participants were asked to perform a lane-change manoeuvre to the left lane after the \textit{test vehicle} crossed a designated point. The driver's inputs were measured through the steering wheel angle, brake, and acceleration pedal positions. 
 All participants were briefed on the required tasks before testing and were allowed to first practice to become familiarized with the simulator. Then, each participant was instructed to conduct 20 consecutive lane change manoeuvres in the same driving scene.

 The identification results for the considered set of drivers are given in TABLE \ref{table:identification}. The average values of driver model parameters are used in the following tests.
 
 \begin{table}[h]
	\begin{center}
		\caption{Identification results of driver parameters }\label{table:identification}
		\begin{tabular}{ccccccc}
			\hline
			\hline
			\textbf{Param.}  & $K_h$& $T_g$ & \textbf{Fit\ $\%$}    \\\hline
	Driver 1& 0.95&0.15&83.14 \\
	Driver 2& 1.15&0.21&86.53 \\
	Driver 3& 1.1&0.14&80.28 \\
	Driver 4& 1.2&0.18&74.53 \\
	Driver 5& 0.98&0.19&80.76 \\
    Average & 1.08&0.17&81.05 \\
    \hline
    \hline
		\end{tabular}
	\end{center}
\end{table}

\subsection{Testing case 1}

\subsubsection{Testing scenario} The testing scenario is set as follows:
\begin{enumerate}[]
\item The driving task is lane keeping and the \textit{ego} vehicle starts from the initial position $(0,0)$.
\item $K_h$ and $T_h$ of the human behavioural model are set as $K_h=1$ and $T_h=0.2s$, respectively.
\item The physical limitations on vehicle actions are set as:
\begin{equation*}
\begin{aligned}
    -30^{\circ}\times \frac{\pi}{180} \leq &\delta_{a,des} (\delta_{h,des}) \leq 30^{\circ}\times \frac{\pi}{180}\\
    -5 \leq &a_{a,des} (a_{h,des}) \leq 5
    \end{aligned}
\end{equation*}
\item The sampling time is $0.05s$. The control horizon is $H_c=3$, while the prediction horizon is $H_p=10$.
\item Other initial setting of the driving scenario is listed in TABLE \ref{table:sce1}.
\begin{table}[h!]
	\begin{center}
		\caption{Environment set-up in the testing }\label{table:sce1}
		\begin{tabular}{|c|c|}
			\hline
			\textbf{Parameter}  & \textbf{Value} \\\hline
			\text{Initial velocity of the \textit{ego}}&  15 $m/s$  \\\hline
			$\text{Initial velocity of the \textit{preceding}}$&12 $m/s$ \\\hline
		\text{Initial distance to the \textit{preceding}}&38 $m$\\\hline
\text{Initial velocity of the \textit{leader}}& 15 $m/s$\\\hline
\text{Initial velocity of the \textit{follower}}&12 $m/s$\\\hline
	\text{Initial distance to the \textit{leader}}&30 $m$\\\hline
		\text{Initial distance to the \textit{follower}}&5 $m$\\\hline
		\end{tabular}
	\end{center}
\end{table}
\item The setting of parameters of APF and DPF is listed in TABLE \ref{table:dpf} where $LC$ denotes \textit{lane changing} and $LK$ denotes \textit{lane keeping}.

\begin{table}[h!]
	\begin{center}
		\caption{Potential field parameters  }\label{table:dpf}
		\begin{tabular}{|c|c|c|c|}
			\hline
			\textbf{Parameter}  & \textbf{Value} &\textbf{Parameter}  & \textbf{Value} \\\hline
			$\alpha_f$& 30 &$\sigma_y$&1\\\hline
	                 $d_c$&0.8 [m] &$v_w$&2 [m]\\\hline
			$\alpha$&10 &$b$&2\\\hline
			$A_f$&2 &$\sigma_x$&4\\\hline
			$a_{max}$&7 $[m/s^2]$& $t_r$&$0.1 [s]$\\\hline
			$t_i$&0.1 [s]&$d_o$& 2 [m]\\\hline
			$[w_f\ w_l\ w_p]$&[0.8\ 0.1\ 0.1] (LC)&$[w_f\ w_l\ w_p]$&[0.1\ 0.2\ 0.7] (LK)\\\hline
			\end{tabular}
	\end{center}
\end{table}

\item The \textit{automation} system degradation is set as
\begin{equation} \label{error_SSfff}
    \delta_{a,f}(k)=\left\{
\begin{array}{rcl}
&\delta_{a,des}(k), &  k < 10\\
&\delta_{a,des}(k)+\frac{0.3\times (k-10)}{20} &  10 \leq k\leq 30\\
&\delta_{a,des}(k)+0.3, & Otherwise
\end{array} \right.\\
\end{equation}
where $k$ is the time step. According to Eq. (\ref{error_SSfff}), it can be seen that a variation is added to the \textit{desire} steering wheel angle after  $t=0.5s$. Testing case 1 with the automation system degradation (Eq. (\ref{error_SSfff})) aims to validate the effectiveness of the proposed approach in lateral direction.
\end{enumerate}

\begin{figure}[h!]
	\begin{center}
		{\includegraphics[width=0.45\textwidth]{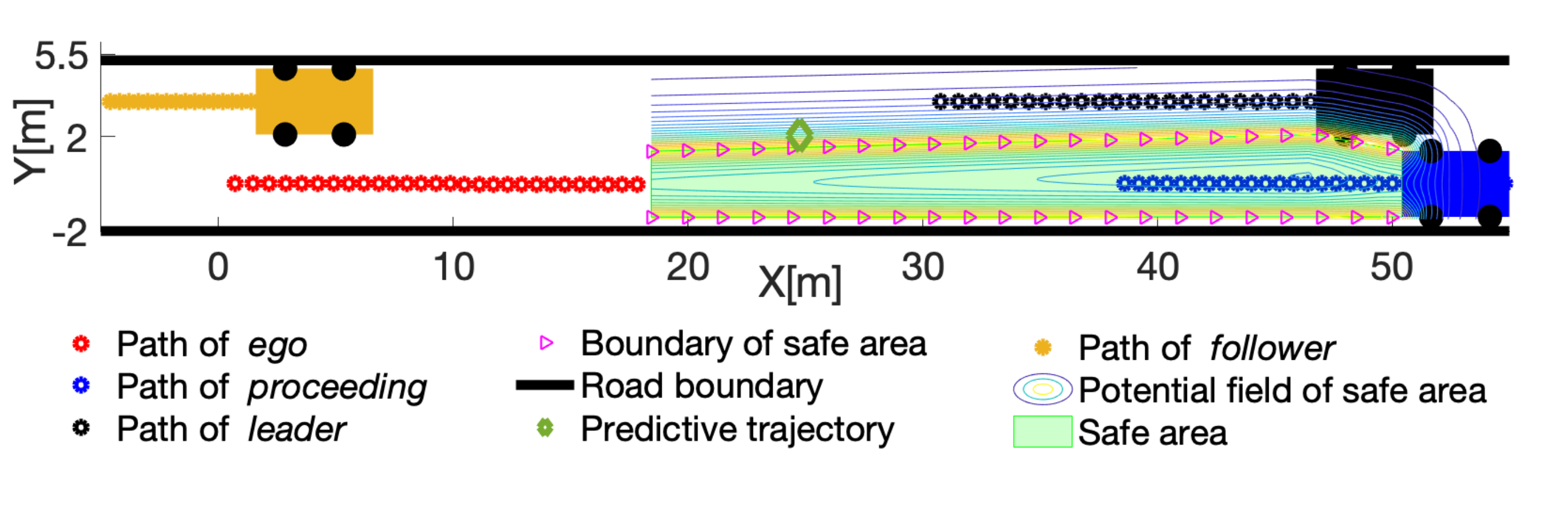}}
		\caption{Results of vehicle's path in the two-lane way}
		\label{fig_case1}
	\end{center}
\end{figure}

\begin{figure}[h!]
	\begin{center}
		{\includegraphics[width=0.48\textwidth]{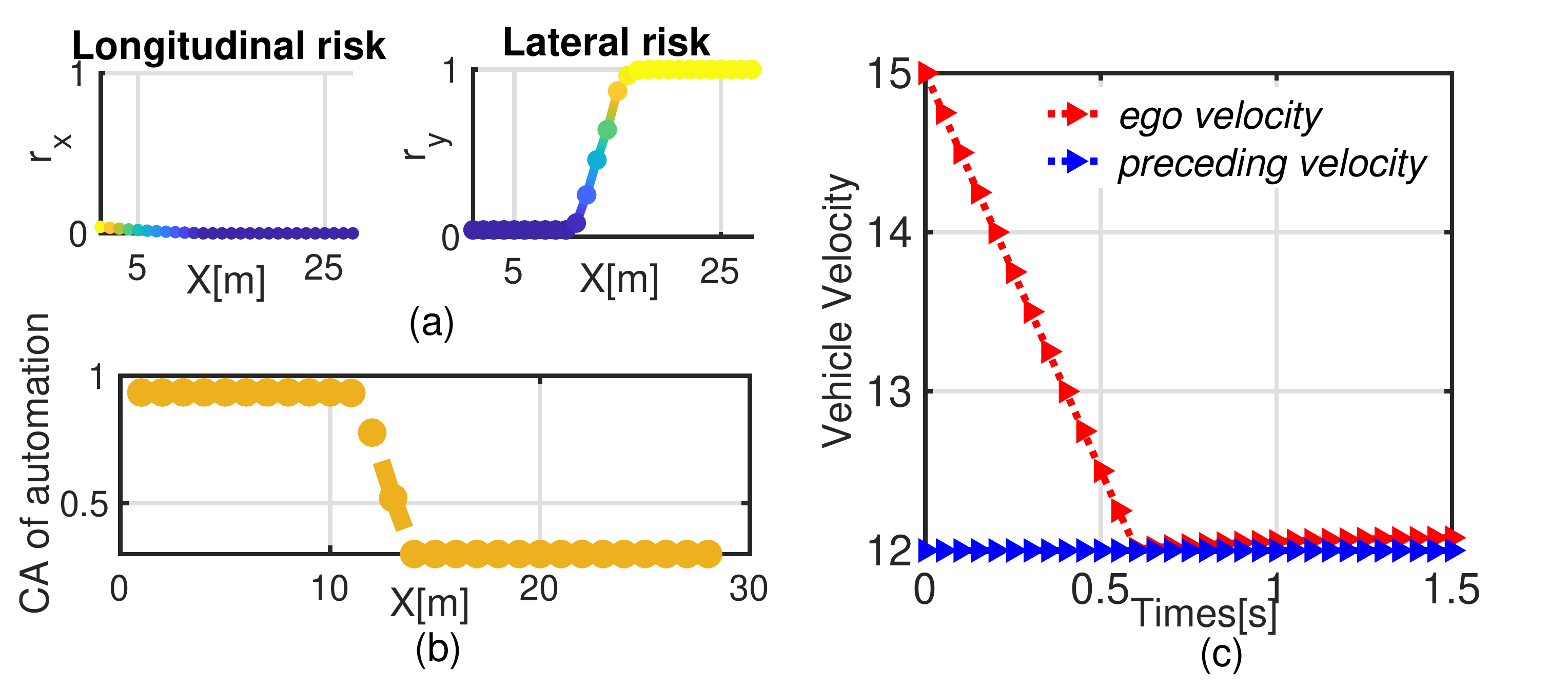}}
		\caption{Risk assessment of \textit{Case} 1}
		\label{fig_case1_rs}
	\end{center}
\end{figure}

\begin{figure}[h!]
    \centering
        \begin{subfigure}[b]{0.24\textwidth}
        \includegraphics[width=\textwidth]{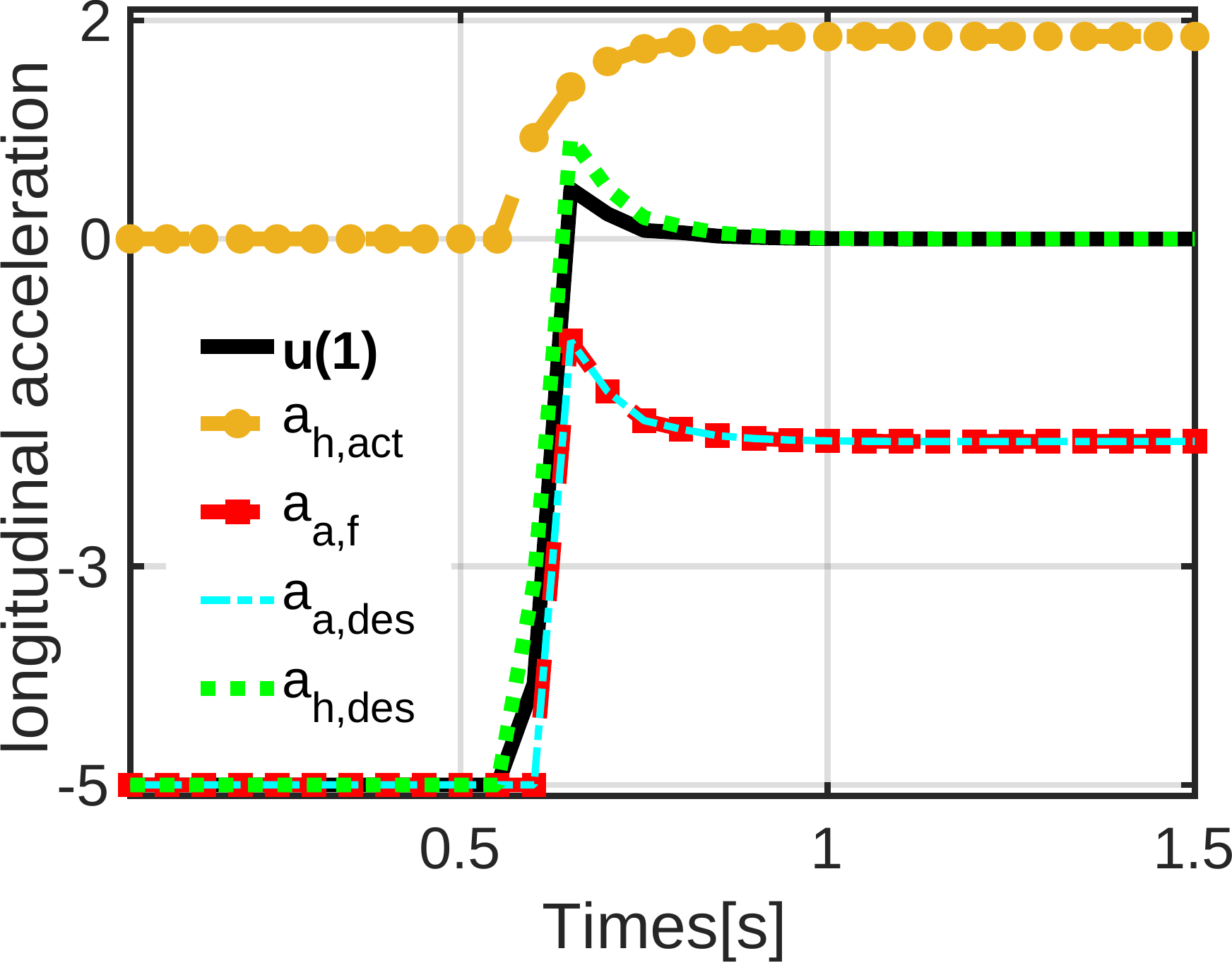}
        \caption{Longitudinal acceleration $a_x$}\label{twolane_propose2}
    \end{subfigure}
        \begin{subfigure}[b]{0.24\textwidth}
        \includegraphics[width=\textwidth]{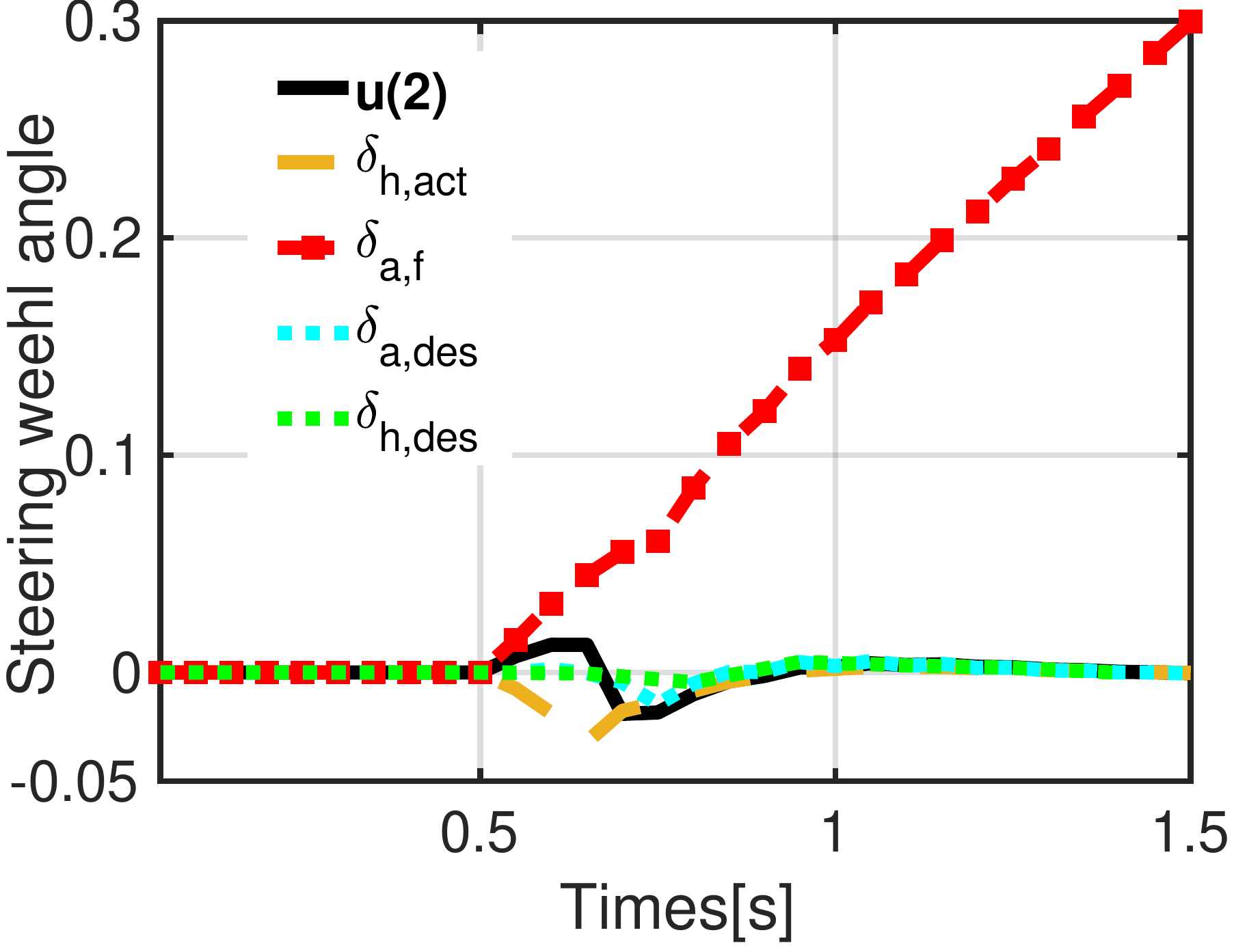}
        \caption{Steering wheel angle $\delta_f$}\label{twolane_driver2}
    \end{subfigure}
    \caption{Control outputs of \textit{Case} 1}\label{fig_case1_u}
\end{figure}

 \subsubsection{Results} Fig. \ref{fig_case1} shows the vehicle motion of \textit{Case} 1 in the two-lane way. It can be seen that in the presence of automation gradation, the \textit{ego} still can complete lane keeping and follow the \textit{preceding}. More specifically, at each time step, CDT can identify the safe driving area for the \textit{ego}, and the boundary of the safe area can be represented by the APF. The predicted trajectory of \textit{ego} can be conducted by the CTRA model and is represented by a {\color{green} green} diamond in Fig. \ref{fig_case1}. In addition, \textit{ego}'s velocity can successfully track that of the \textit{preceding}, as shown in Fig.  \ref{fig_case1_rs}c. Fig. \ref{fig_case1_rs}a shows the driving risk over time. 
 The estimated driving risk indicates the occurrence of performance degradation, especially in the lateral direction. The corresponding maximum control authority of \textit{automation} ($\alpha_a$) is determined by the FIS and shown in Fig. \ref{fig_case1_rs}b. $\alpha_a$ is varied based on the severity of the performance degradation. Fig. \ref{fig_case1_u} represents the control outputs of the \textit{automation} and \textit{human driver}. In the normal operation condition ($t<0.5s$), $u_{a,f}=u_{a,des}$, and the {\color{red} red} solid line and {\color{cyan} cyan} dash line coincide with each other. Since the \textit{human driver} and \textit{automation} share the same objective, i.e., $u_{h,des}\approx u_{a,des}$, the \textit{automation} can deal with the driving task ($\mathbf{u}\approx u_{a,des}$) and the \textit{human driver} does not need to exert additional control effort ($u_{h,act}\rightarrow 0$). Once the occurrence of performance degradation is identified by the assessment of driving risk, the \textit{actual} control output of \textit{automaton} is managed by $\alpha_a$ through Eq. (\ref{actual_u}). As shown in Fig. \ref{fig_case1_u}b, the value of $\delta_{a,act}$ is much smaller than $\delta_{a,f}$ and no larger than $\delta_{a,des}$. In this way, the impact of $\delta_{a,f}$ on vehicle motion is limited, and the \textit{human driver} is able to compensate for the performance degradation by exerting $\delta_{h,act}$. By the contrast, there is no change in longitudinal acceleration; then, we have $a_{a,f}=a_{a,des}$. In addition, as shown in Fig. \ref{fig_case1_u}a after $t=0.6s$, the \textit{human driver} has a priority over \textit{automation} when they do not have same objective ($u_{h,des} \neq u_{a,des}$). The proposed shared control algorithm allows the driver to own the final authority over the automation. The actual control output of the shared control system $\mathbf{u}$  highly depends on the human driver's desire. 

\subsection{Testing case 2}
In \textit{Case} 2, we examine the effectiveness of the proposed approach on a curvy road. The road radius is $R=60m$. The initial velocity of the \textit{ego} and the \textit{preceding} is $12 m/s$. The initial distance is $15m$. The parameters of simulation setting are as the same as those of  \textit{Case} 1. The \textit{automation} system degradation is set as

\begin{equation} \label{error_fff}
    a_{a,f}(k)=\left\{
\begin{array}{rcl}
&a_{a,des}(k), &  k \leq 5\\
&a_{a,des}(k)+\frac{3 \times (k-5) }{5} & 5 < k\leq 10\\
&a_{a,des}(k)+3, & Otherwise
\end{array} \right.\\
\end{equation}
From Eq. (\ref{error_fff}), it can be seen that after $k>5$, the value of longitudinal acceleration is gradually increased. After $k>10$, a permanent failure occurs and $a_{a,f}$ is always equal to $a_{a,des}+3$. It should be noted that $a_{a,f}$ also satisfies $-5 \leq a_{a,f} \leq 5$. Case 2 mainly shows the advantage of the proposed approach in the condition of \textit{automaton} system degradation in the longitudinal direction.

\begin{figure}[h!]
	\begin{center}
		{\includegraphics[width=0.48\textwidth]{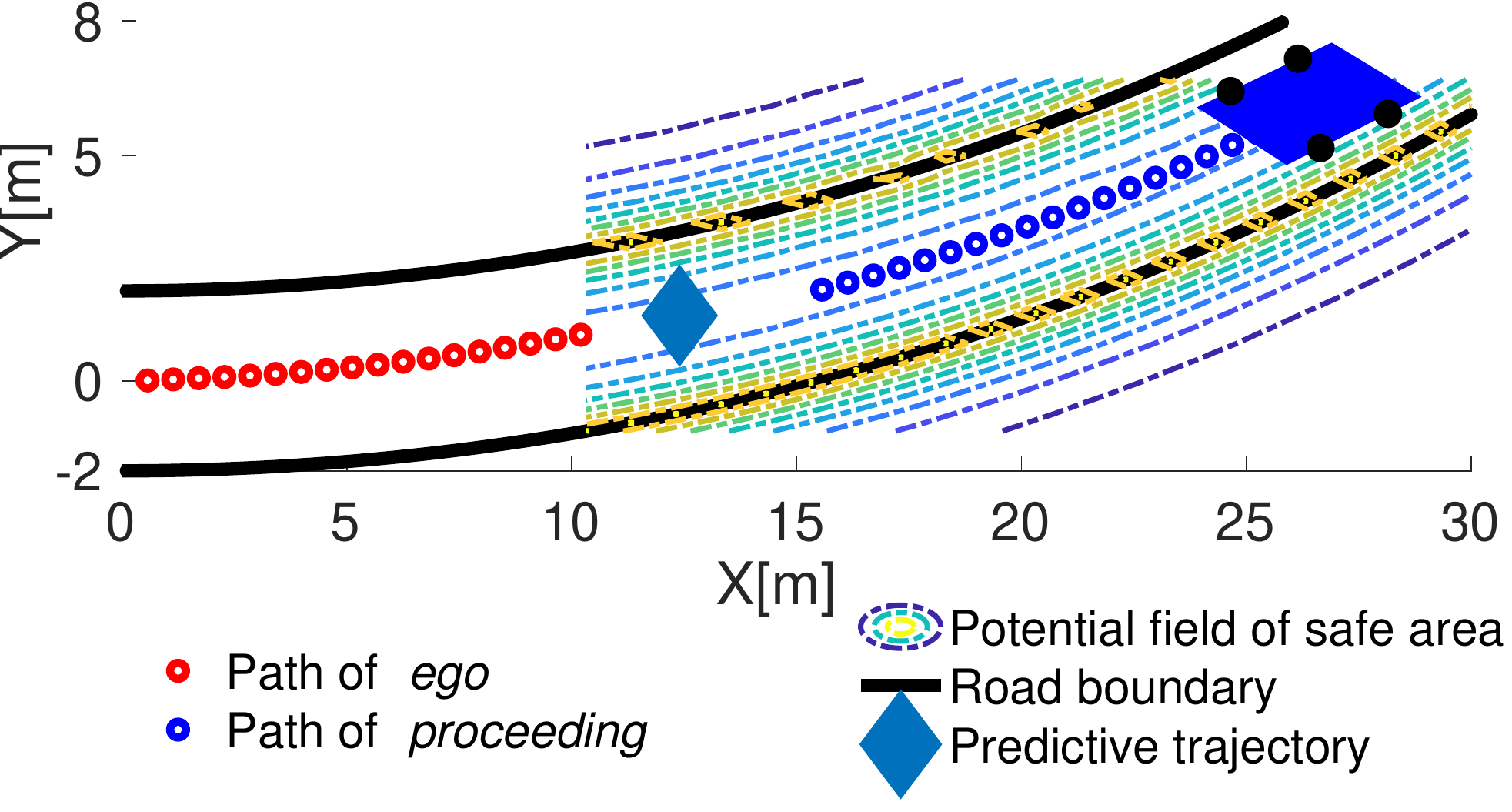}}
		\caption{Results of vehicle's path on the curved road}
		\label{fig_case2}
	\end{center}
\end{figure}

\begin{figure}[h!]
	\begin{center}
		{\includegraphics[width=0.48\textwidth]{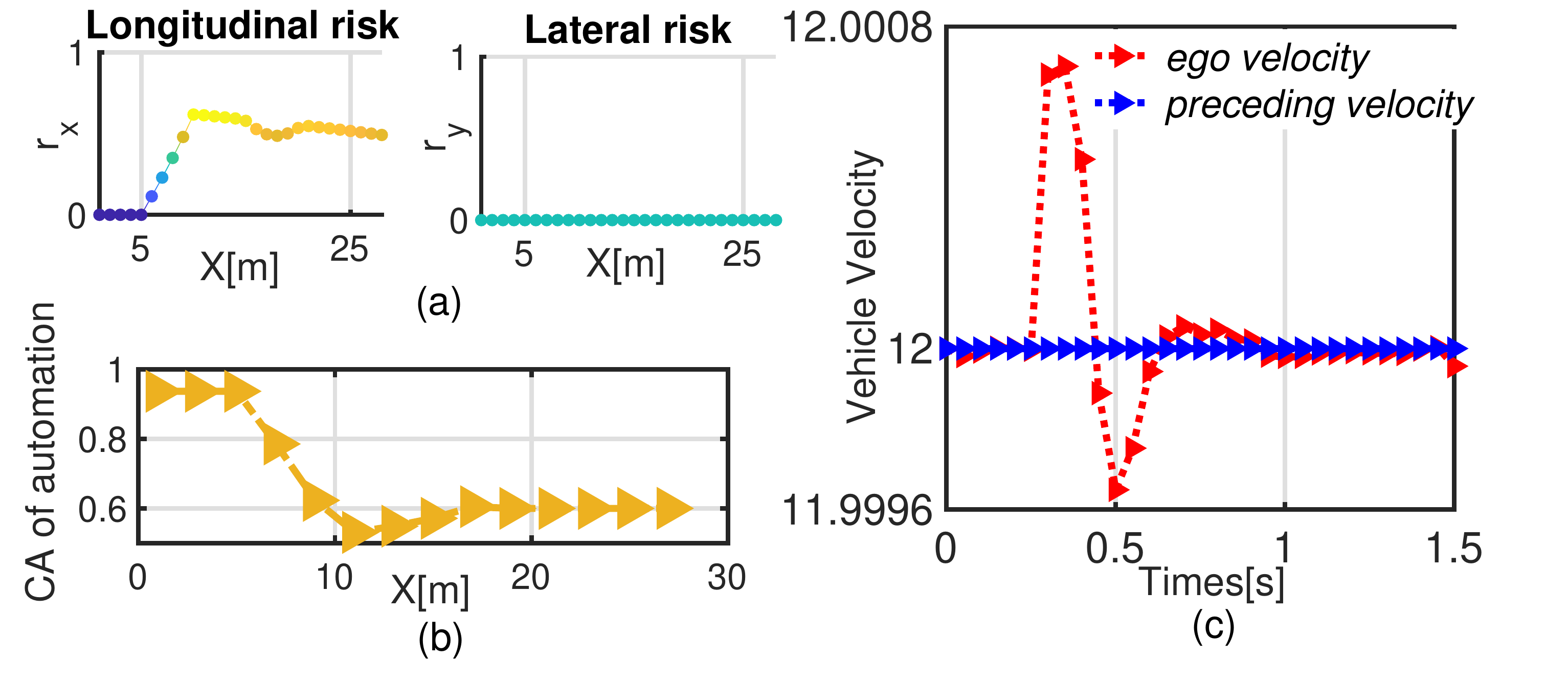}}
		\caption{Risk assessment of \textit{Case} 2}
		\label{fig_case2_rs}
	\end{center}
\end{figure}

\begin{figure}[h!]
    \centering
        \begin{subfigure}[b]{0.24\textwidth}
        \includegraphics[width=\textwidth]{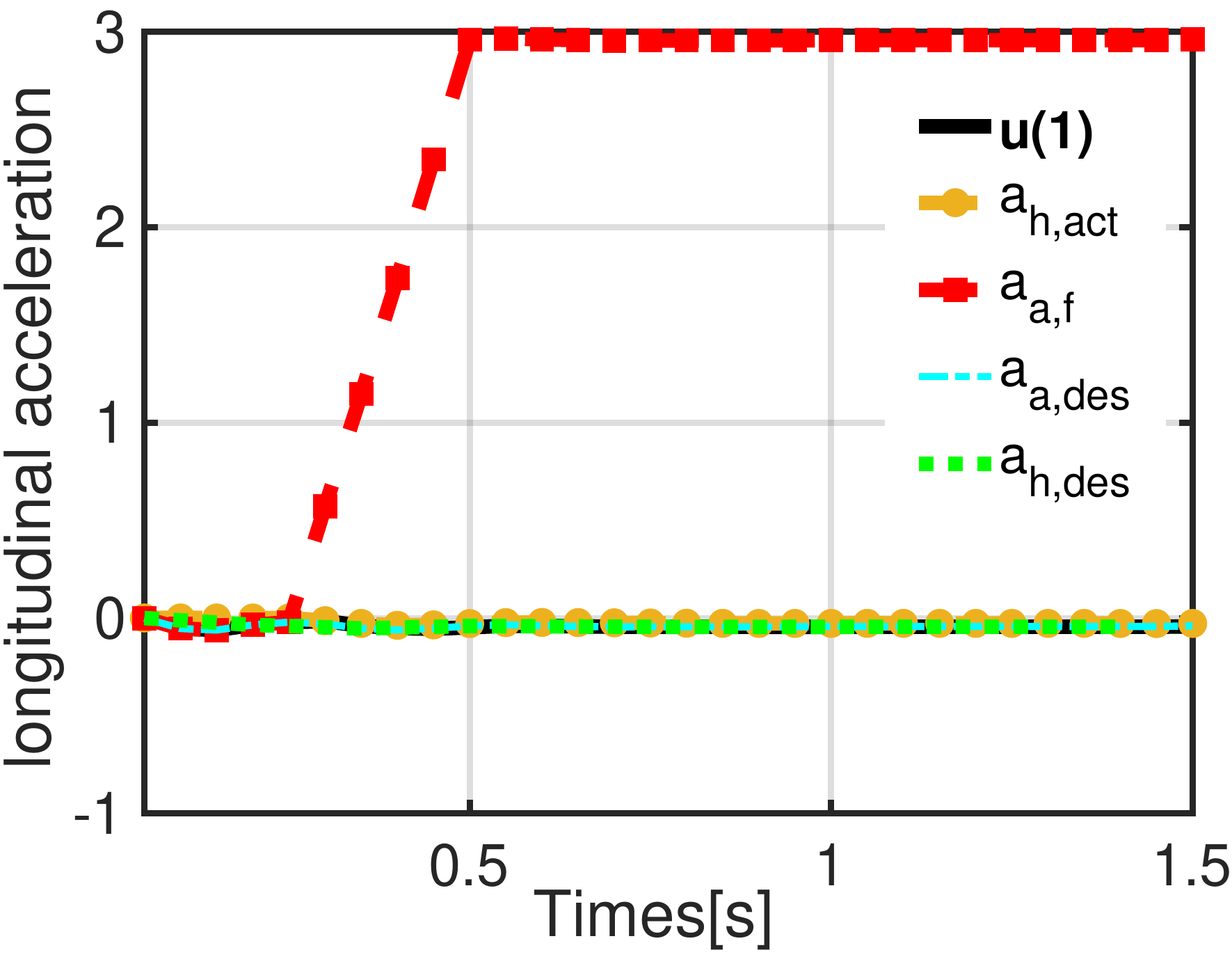}
        \caption{Longitudinal acceleration $a_x$}\label{twolane_propose3}
    \end{subfigure}
        \begin{subfigure}[b]{0.24\textwidth}
        \includegraphics[width=\textwidth]{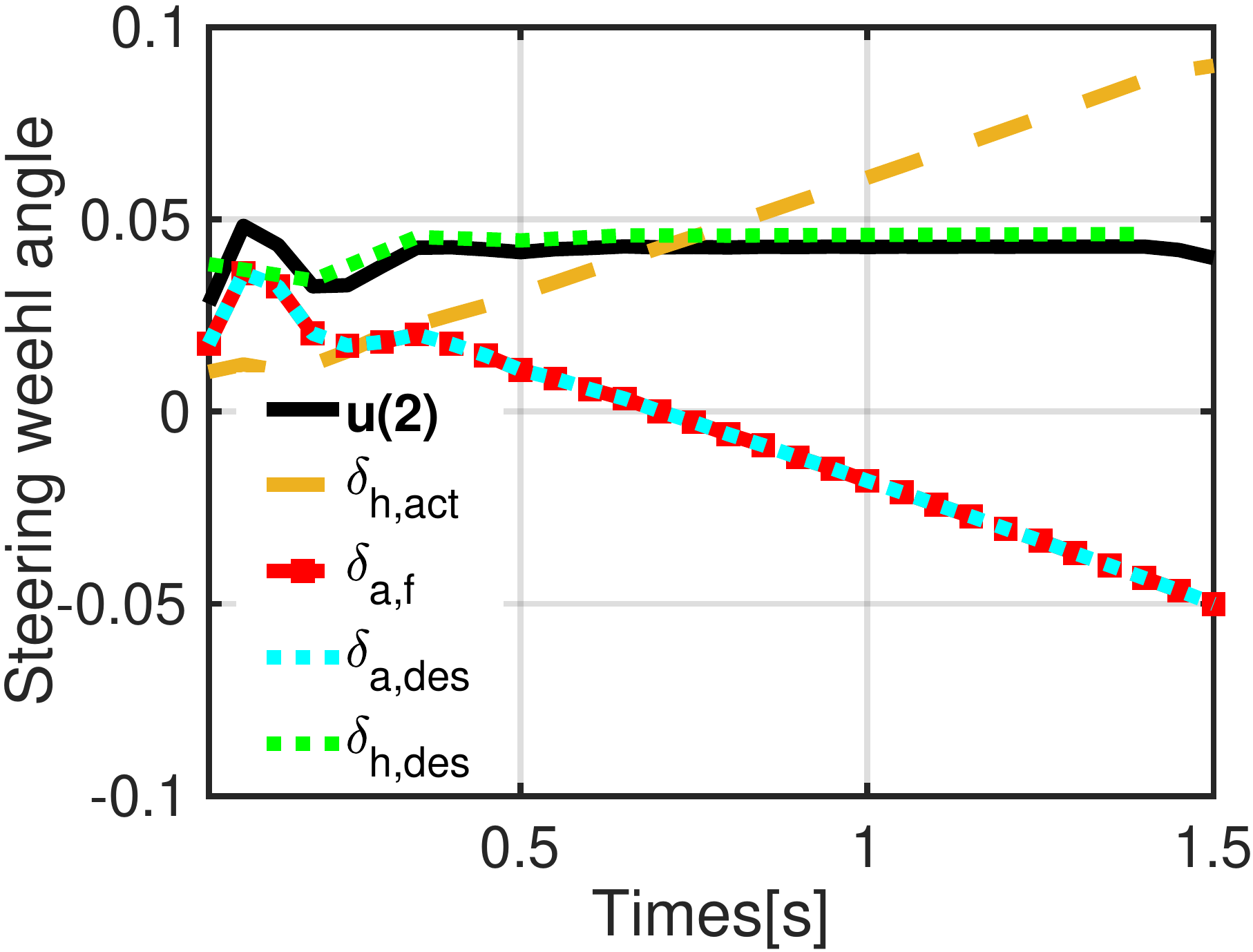}
        \caption{Steering wheel angle $\delta_f$}\label{twolane_driver3}
    \end{subfigure}
    \caption{Control outputs of \textit{Case} 2}\label{fig_case2_u}
\end{figure}

Fig. \ref{fig_case2} indicates that \textit{ego} is able to perform its driving task through the control input $\mathbf{u}$ given by the \textit{automation} and \textit{human driver}. Furthermore, only a small vibration  of \textit{ego}'s velocity occurs at the initial phase of the performance degradation, as shown in Fig. \ref{fig_case2_rs}c. Fig. \ref{fig_case2_rs}a shows that the longitudinal risk is increased with the severity of degradation on the longitudinal acceleration $a_{a,f}$. The corresponding control authority of \textit{automation} varies according to the driving risk. Fig. \ref{fig_case2_u} shows the allocation of control inputs in the proposed shared control system. As $a_{a,f}\gg a_{a,des}$, to reduce the influence of performance degradation, the value of actual longitudinal acceleration of \textit{automaton} ($a_{a,act}$) is bounded by $\alpha \times u_{a,des}$.  In this way, the \textit{human driver} is able to compensate to the required control input. Therefore, the whole control vector, which is composed of $u_{a,act}$ and $u_{h,act}$, is imported to the vehicle system, guaranteeing the safety of the driving task.

\section{Conclusion}\label{con}

In this paper, a novel shared driving control algorithm was proposed for automated vehicles under automation degradation. The driving risks, including the longitudinal risk and lateral risk, are assessed to identify the severity of performance degradation based on the vehicle's trajectory prediction. The control authority of \textit{automation} can be adaptively updated in real time according to the assessed driving risks. Once automation degradation is detected, the value of the control authority allocated to the \textit{automation} is bounded. This allows the \textit{human driver} to compensate for the performance degradation by providing the additional required control effort and thus ensure the driving safety. The feasibility and effectiveness of the proposed shared control algorithm were validated in two typical scenarios of automated driving. Future works will be focusing on further improvement of the proposed approach and its implementation in a real vehicle platform.


\section*{Acknowledgments}

This work was supported in part by the SUG-NAP Grant (No. M4082268.050) of Nanyang Technological University, Singapore, and the A*STAR Grant (No. 1922500046), Singapore.

\bibliographystyle{IEEEtran}
\bibliography{LSTM_main.bib}

\begin{thebibliography}{10}
\providecommand{\url}[1]{#1}
\csname url@samestyle\endcsname
\providecommand{\newblock}{\relax}
\providecommand{\bibinfo}[2]{#2}
\providecommand{\BIBentrySTDinterwordspacing}{\spaceskip=0pt\relax}
\providecommand{\BIBentryALTinterwordstretchfactor}{4}
\providecommand{\BIBentryALTinterwordspacing}{\spaceskip=\fontdimen2\font plus
\BIBentryALTinterwordstretchfactor\fontdimen3\font minus
  \fontdimen4\font\relax}
\providecommand{\BIBforeignlanguage}[2]{{%
\expandafter\ifx\csname l@#1\endcsname\relax
\typeout{** WARNING: IEEEtran.bst: No hyphenation pattern has been}%
\typeout{** loaded for the language `#1'. Using the pattern for}%
\typeout{** the default language instead.}%
\else
\language=\csname l@#1\endcsname
\fi
#2}}
\providecommand{\BIBdecl}{\relax}
\BIBdecl

\bibitem{xu2019integrated}
C.~Xu, W.~Zhao, and C.~Wang, ``An integrated threat assessment algorithm for
  decision-making of autonomous driving vehicles,'' \emph{IEEE Transactions on
  Intelligent Transportation Systems}, 2019.

\bibitem{hang2020human}
P.~Hang, C.~Lv, Y.~Xing, C.~Huang, and Z.~Hu, ``Human-like decision making for
  autonomous driving: A noncooperative game theoretic approach,'' \emph{IEEE
  Transactions on Intelligent Transportation Systems}, 2020.

\bibitem{hoc2001towards}
J.-M. Hoc, ``Towards a cognitive approach to human--machine cooperation in
  dynamic situations,'' \emph{International journal of human-computer studies},
  vol.~54, no.~4, pp. 509--540, 2001.

\bibitem{lv2017analysis}
C.~Lv, D.~Cao, Y.~Zhao, D.~J. Auger, M.~Sullman, H.~Wang, L.~M. Dutka,
  L.~Skrypchuk, and A.~Mouzakitis, ``Analysis of autopilot disengagements
  occurring during autonomous vehicle testing,'' \emph{IEEE/CAA Journal of
  Automatica Sinica}, vol.~5, no.~1, pp. 58--68, 2017.

\bibitem{wang2020decision}
W.~Wang, X.~Na, D.~Cao, J.~Gong, J.~Xi, Y.~Xing, and F.-Y. Wang,
  ``Decision-making in driver-automation shared control: A review and
  perspectives,'' \emph{IEEE/CAA Journal of Automatica Sinica}, vol.~7, no.~5,
  pp. 1289--1307, 2020.

\bibitem{marcano2020review}
M.~Marcano, S.~D{\'\i}az, J.~P{\'e}rez, and E.~Irigoyen, ``A review of shared
  control for automated vehicles: Theory and applications,'' \emph{IEEE
  Transactions on Human-Machine Systems}, 2020.

\bibitem{huang2020reference}
C.~Huang, C.~Lv, F.~Naghdy, and H.~Du, ``Reference-free approach for mitigating
  human--machine conflicts in shared control of automated vehicles,'' \emph{IET
  Control Theory \& Applications}, vol.~14, no.~18, pp. 2752--2763, 2020.

\bibitem{flemisch2012towards}
F.~Flemisch, M.~Heesen, T.~Hesse, J.~Kelsch, A.~Schieben, and J.~Beller,
  ``Towards a dynamic balance between humans and automation: authority,
  ability, responsibility and control in shared and cooperative control
  situations,'' \emph{Cognition, Technology \& Work}, vol.~14, no.~1, pp.
  3--18, 2012.

\bibitem{benloucif2017new}
M.~A. Benloucif, A.-T. Nguyen, C.~Sentouh, and J.-C. Popieul, ``A new scheme
  for haptic shared lateral control in highway driving using trajectory
  planning,'' \emph{IFAC-PapersOnLine}, vol.~50, no.~1, pp. 13\,834--13\,840,
  2017.

\bibitem{nguyen2016driver}
A.-T. Nguyen, C.~Sentouh, and J.-C. Popieul, ``Driver-automation cooperative
  approach for shared steering control under multiple system constraints:
  Design and experiments,'' \emph{IEEE Transactions on Industrial Electronics},
  vol.~64, no.~5, pp. 3819--3830, 2016.

\bibitem{vasconez2019design}
J.~P. Vasconez, D.~Carvajal, and F.~A. Cheein, ``On the design of a
  human--robot interaction strategy for commercial vehicle driving based on
  human cognitive parameters,'' \emph{Advances in Mechanical Engineering},
  vol.~11, no.~7, p. 1687814019862715, 2019.

\bibitem{wada2017shared}
T.~Wada and R.~Kondo, ``Shared authority mode: Connecting automated and manual
  driving for smooth authority transfer,'' \emph{Proceedings of FAST-zero},
  2017.

\bibitem{erden2010human}
M.~S. Erden and T.~Tomiyama, ``Human-intent detection and physically
  interactive control of a robot without force sensors,'' \emph{IEEE
  Transactions on Robotics}, vol.~26, no.~2, pp. 370--382, 2010.

\bibitem{xing2019driver}
Y.~Xing, C.~Lv, H.~Wang, H.~Wang, Y.~Ai, D.~Cao, E.~Velenis, and F.-Y. Wang,
  ``Driver lane change intention inference for intelligent vehicles: Framework,
  survey, and challenges,'' \emph{IEEE Transactions on Vehicular Technology},
  vol.~68, no.~5, pp. 4377--4390, 2019.

\bibitem{huang2018delta}
C.~Huang, F.~Naghdy, and H.~Du, ``Delta operator-based model predictive control
  with fault compensation for steer-by-wire systems,'' \emph{IEEE Transactions
  on Systems, Man, and Cybernetics: Systems}, 2018.

\bibitem{8114322}
C.~{Huang}, F.~{Naghdy}, and H.~{Du}, ``Fault tolerant sliding mode predictive
  control for uncertain steer-by-wire system,'' \emph{IEEE Transactions on
  Cybernetics}, vol.~49, no.~1, pp. 261--272, 2019.

\bibitem{favaro2018autonomous}
F.~Favar{\`o}, S.~Eurich, and N.~Nader, ``Autonomous vehicles’
  disengagements: Trends, triggers, and regulatory limitations,''
  \emph{Accident Analysis \& Prevention}, vol. 110, pp. 136--148, 2018.

\bibitem{dixit2016autonomous}
V.~V. Dixit, S.~Chand, and D.~J. Nair, ``Autonomous vehicles: disengagements,
  accidents and reaction times,'' \emph{PLoS one}, vol.~11, no.~12, p.
  e0168054, 2016.

\bibitem{waschl2019control}
H.~Waschl, I.~Kolmanovsky, and F.~Willems, ``Control strategies for advanced
  driver assistance systems and autonomous driving functions,''
  \emph{Switzerland: Springer International}, 2019.

\bibitem{lefevre2014survey}
S.~Lef{\`e}vre, D.~Vasquez, and C.~Laugier, ``A survey on motion prediction and
  risk assessment for intelligent vehicles,'' \emph{ROBOMECH journal}, vol.~1,
  no.~1, pp. 1--14, 2014.

\bibitem{nodine2019indicators}
E.~E. Nodine, D.~L. Fisher, G.~Golembiewski, C.~Armstrong, A.~H. Lam, M.~A.
  Jeffers, W.~G. Najm, S.~Miller, S.~Jackson, N.~Kehoe \emph{et~al.},
  ``Indicators of driver adaptation to forward collision warnings: A
  naturalistic driving evaluation,'' United States. Department of
  Transportation. National Highway Traffic Safety~…, Tech. Rep., 2019.

\bibitem{wang2016driving}
J.~Wang, J.~Wu, X.~Zheng, D.~Ni, and K.~Li, ``Driving safety field theory
  modeling and its application in pre-collision warning system,''
  \emph{Transportation research part C: emerging technologies}, vol.~72, pp.
  306--324, 2016.

\bibitem{schubert2011empirical}
R.~Schubert, C.~Adam, M.~Obst, N.~Mattern, V.~Leonhardt, and G.~Wanielik,
  ``Empirical evaluation of vehicular models for ego motion estimation,'' in
  \emph{2011 IEEE intelligent vehicles symposium (IV)}.\hskip 1em plus 0.5em
  minus 0.4em\relax IEEE, 2011, pp. 534--539.

\bibitem{benloucif2019cooperative}
A.~Benloucif, A.-T. Nguyen, C.~Sentouh, and J.-C. Popieul, ``Cooperative
  trajectory planning for haptic shared control between driver and automation
  in highway driving,'' \emph{IEEE Transactions on Industrial Electronics},
  vol.~66, no.~12, pp. 9846--9857, 2019.

\bibitem{fan2020improved}
X.~Fan, Y.~Guo, H.~Liu, B.~Wei, and W.~Lyu, ``Improved artificial potential
  field method applied for auv path planning,'' \emph{Mathematical Problems in
  Engineering}, vol. 2020, 2020.

\bibitem{tavana2013fuzzy}
M.~Tavana, F.~Azizi, F.~Azizi, and M.~Behzadian, ``A fuzzy inference system
  with application to player selection and team formation in multi-player
  sports,'' \emph{Sport Management Review}, vol.~16, no.~1, pp. 97--110, 2013.

\bibitem{ji2018shared}
X.~Ji, K.~Yang, X.~Na, C.~Lv, and Y.~Liu, ``Shared steering torque control for
  lane change assistance: A stochastic game-theoretic approach,'' \emph{IEEE
  Transactions on Industrial Electronics}, vol.~66, no.~4, pp. 3093--3105,
  2018.

\bibitem{na2019modelling}
X.~Na and D.~J. Cole, ``Modelling of a human driver's interaction with vehicle
  automated steering using cooperative game theory,'' \emph{IEEE/CAA Journal of
  Automatica Sinica}, vol.~6, no.~5, pp. 1095--1107, 2019.

\end{thebibliography}
\vspace{-2cm}

\end{document}